\documentclass[prb,twocolumn,amsmath,amssymb,floatfix,footinbib,bibnotes]{revtex4}

\pdfoutput=1

\usepackage[colorlinks=true,citecolor=blue,linkcolor=magenta]{hyperref}
\usepackage{amsmath}
\usepackage{graphicx}
\usepackage{dsfont}
\usepackage{changepage}
\usepackage{appendix}
\usepackage[T1]{fontenc}
\usepackage[latin1]{inputenc}

\newcommand{\be}{\begin{equation}}
\newcommand{\ee}{\end{equation}}

\def \ua{{\uparrow}}
\def \da{{\downarrow}}
\def \be{\begin{equation}}
\def \ee{\end{equation}}
\def \ba{\begin{array}}
\def \ea{\end{array}}
\def \bea{\begin{eqnarray}}
\def \eea{\end{eqnarray}}
\def \nn{\nonumber}
\def \cH{{\cal{H}}}

\def \tu{{\tilde{u}}}
\def \tlr{{\tilde{r}}}
\def \W{{\Omega}}
\def \e{{\epsilon}}

\def \a{{\alpha}}
\def \t{{\theta}}
\def \b{{\beta}}
\def \g{{\gamma}}
\def \D{{\Delta}}
\def \d{{\delta}}
\def \w{{\omega}}
\def \s{{\sigma}}
\def \f{{\varphi}}

\def \e{{\epsilon}}

\def \yd{^\dagger}
\def \av#1{{\langle#1\rangle}}

\def \W{{\omega}}

\def \beas{\begin{eqnarray*}}
\def \eeas{\end{eqnarray*}}

\def \mrm{\mathrm}

\newcounter{indice}

\begin{document}
\title{Non-local Order in Elongated Dipolar Gases}

\author{J. Ruhman$^1$ , E. G. Dalla Torre$^{1,2}$ , S. D. Huber$^1$  and E. Altman$^1$ \\
{$^1$\small \em Department of Condensed Matter Physics, Weizmann Institute of Science, Rehovot 76100, Israel}\\
{$^2$\small \em Department of Physics, Harvard University, Cambridge MA 02138}}
\begin{abstract}
Dipolar particles in an elongated trap are expected to undergo a quantum phase transition from a linear to a zigzag structure with decreasing transverse confinement. We derive the low energy effective theory of the transition showing that in presence of quantum fluctuations the zigzag phase can be characterized by a long ranged string order, while the local Ising correlations decay as a power law. This is also confirmed using density matrix renormalization group (DMRG) calculations on a microscopic model. The non local order in the bulk gives rise to zero energy states localized at the interface between the ordered and disordered phases. Such an interface naturally arises when the particles are subject to a weak harmonic confinement along the tube axis. We compute the signature of the edge states in the single particle tunneling spectra pointing to differences between a system with Bosonic versus fermionic particles. Finally we asses the magnitude of the relevant quantum fluctuations in realistic systems of dipolar particles, including ultracold polar molecules as well as alkali atoms weakly dressed by a Rydberg excitation.
\end{abstract}
\maketitle
%\numberwithin{equation}{subsection}

\section{Introduction}
The realization of ultracold dipolar gases opens new directions for investigation of quantum many body physics. Relevant systems currently under investigation include degenerate gases of atoms with large magnetic dipole moments, such as Cr\cite{Lahaye2007b} or of heteronuclear molecules with a large permanent electric dipole.\cite{Ni2008} Another promising proposal is to use degenerate Alkali atoms which are weakly  dressed with a Rydberg excitation by optical pumping.\cite{Pupillo2010}

The long range dipolar interactions in these systems can be strong enough to drive interesting structural phase transitions. But, not so strong as to make the kinetic energy negligible. The balance between kinetic and interaction terms results in strong quantum fluctuations, which provide a fertile ground to formation of novel phases.\cite{HIprl,Cooper2009}

\begin{figure}
\centering
\includegraphics[width=8.6cm,height=3.6cm]{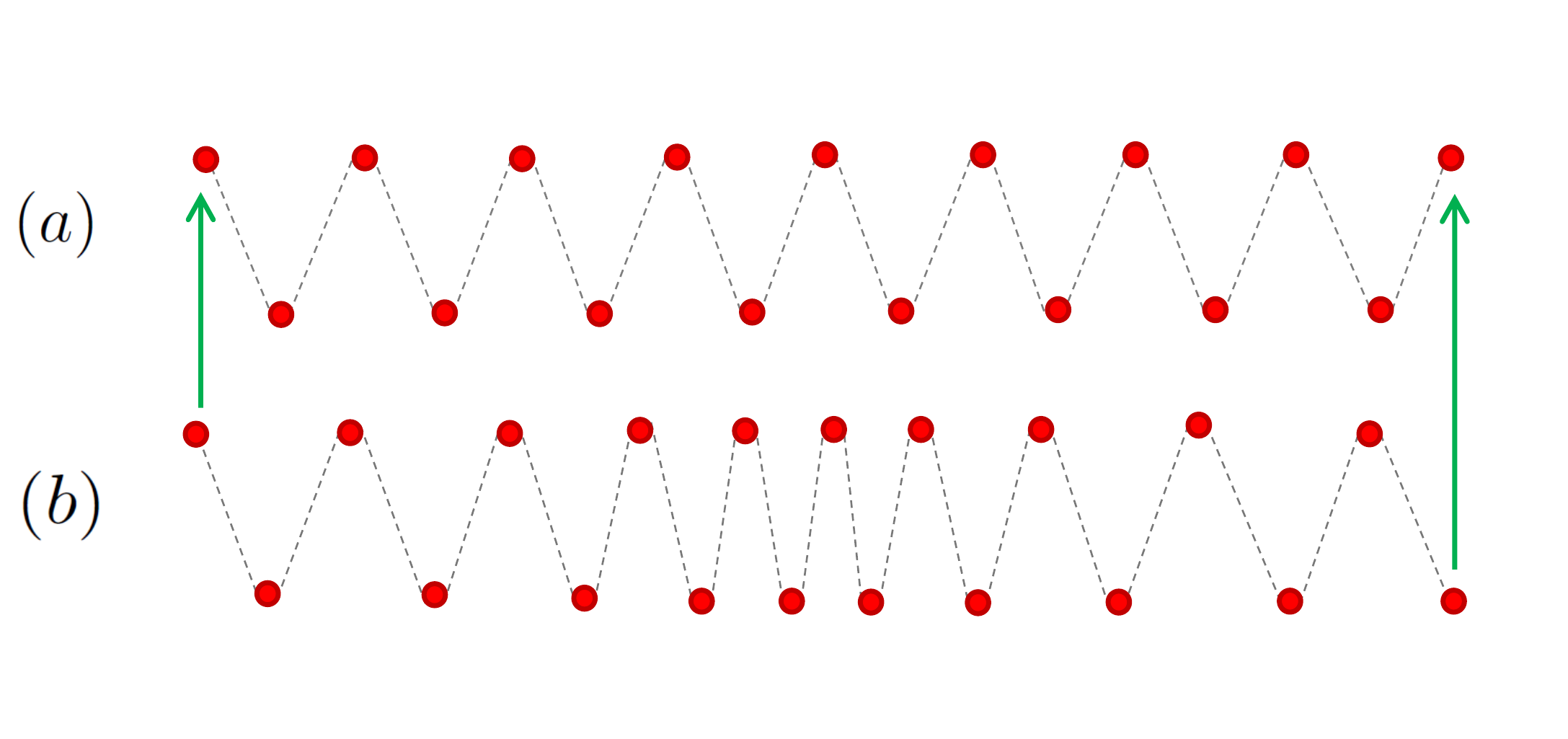}
\caption{(a) One of the two broken symmetry zigzag states of the classical system. (b) Longitudinal quantum fluctuations affect a continuous distortion from one zigzag configuration to the other, which restores the $\mathds{Z}_2$ symmetry of the ground state in the bulk} \label{fig:scheme}
\end{figure}

In this paper we shall specifically consider the zigzag instability of a chain of repulsive particles in an elongated trap, shown schematically in Fig. \ref{fig:scheme}. The dipoles are assumed to be polarized by an external electric field perpendicular to the trap axis. When the transverse confinement is lowered below a critical value, the repulsive interactions between dipoles overcomes the confinement, leading to a staggered distortion of the chain. A classical analysis along these lines\cite{Fishman2008} explains the zigzag distortion observed in chains of trapped ions.\cite{Birkl1992} In the classical description of either the coulomb or dipolar crystal, the zigzag distortion is associated with breaking of the $\mathds{Z}_2$ reflection symmetry about the midplane of the trap. However, because the dipolar interactions are much weaker, the effect of quantum fluctuations in the particle positions is greatly enhanced compared to the ions. We shall see that this leads to an interesting change of the zigzag phase and phase transition.

One effect of quantum fluctuations is to allow single particle tunneling between an up and down displacement in the zigzag. It is convenient to think of distortion of a particle, up or down as an Ising spin variable. The tunneling then acts to disorder the zigzag in the same way that a transverse field acts to disorder a one dimensional Ising model inducing transitions between the up and down state of the spin. This class of fluctuations and a mapping to an Ising model were recently discussed.\cite{Shimshoni2011,Shimshoni2011a}

Another mode of quantum fluctuation, which has not been considered in this context so far, is the axial motion of the dipolar particles. In one dimension these longitudinal fluctuations lead to disordering of the crystal and at the same time they restore the $\mathds{Z}_2$ symmetry of the staggered configuration. What is then the fate of the phase transition and of the broken symmetry phase in the presence of the fluctuations?

To answer this question we derive a long wavelength theory, which describes both the gapless longitudinal fluctuations and the transverse distortion. Let us briefly summarize the main results of this analysis. First, we show that the zigzag state remains a distinct phase in presence of the fluctuations, however the zigzag order becomes non-local and is described by a string order-parameter. The longitudinal density fluctuations (breathing modes of the zigzag) remain gapless in the zigzag phase, while single particle excitations induce domain walls in the string order and are therefore gapped. We further show that the presence of string order in the bulk implies the existence of a zero energy single-particle state localized at each edge. These states are represented by localized Majorana modes in the low energy theory. We compute the local tunneling spectra into the edge and identify the signatures of the localized zero energy states. Finally we point to interesting differences in these signatures depending on whether the dipolar particles are of a Bosonic or a fermionic species.

Before proceeding we note previous analysis done in the context of electronic quantum wires,\cite{Meyer2007,Meyer2009,Sitte2009} which also identify a quantum phase transition, of the same universality class as we consider here, into a zigzag chain. While the non-local nature of the order parameter is implicit in these studies, this non-locality has not been emphasized, and in particular its implications on the edge structure have not been discussed. We also note earlier studies on Bosonic\cite{Orignac1998} and Fermionic\cite{Starykh2000} ladders, where a state similar to the zigzag is identified as an out of phase charge density wave between the two legs of the ladder.

The rest of the paper is organized as follows. In section \ref{sec:H} we identify the important scales in the problem of dipolar particles in an elongated trap and derive an effective Hamiltonian of the particles in first quantized form. A quantum mean field approximation is then formulated in section \ref{sec:mf} in order to estimate the transition point and asses the importance of quantum fluctuations within the interaction range relevant to ultra-cold dipolar molecules or Rydberg atoms. In section \ref{sec:ft} we derive an effective long-wavelength theory of the Zigzag transition starting from the microscopic Hamiltonian of the dipolar chain. Using the long-wave description we discuss the nature of the critical point and the non local order parameter associated with the zigzag phase. The low energy edge states imposed by the non-local order in the bulk are discussed in section \ref{sec:edge}. In particular we predict direct signatures of the zero energy edge states in local tunneling spectra. In section \ref{sec:dmrg} we demonstrate the key properties of the zigzag chain, in particular the presence of string order, in a numerical calculation using the density matrix renormalization group (DMRG) method. Finally section \ref{sec:summary} provides a summary and discussion of the results.

\section{Model Hamiltonian}\label{sec:H}
We consider a dipolar quantum gas tightly confined in a tube-shaped trapping potential. The particles of the gas may be either fermionic or Bosonic and we assume that their dipoles are polarized by a large electric field (or magnetic field in the case of atoms with a large magnetic dipole moment) in the direction $\hat{\bf z}$ perpendicular to the tube axis. Such a setup is described by the Hamiltonian
\begin{align}
H=&\sum_i\left({{P}_i^2\over 2m} +{m \omega_\bot ^2 \over 2}r_i ^2\right) \label{Full_Hamiltonian}\\+\,&d^2\sum_{j> i}\left({1\over |\textbf{R}_i - \textbf{R}_j|^3}-3{\left(( \textbf{R}_i- \textbf{R}_j )\cdot {\hat{\bf z}}\right)^2\over |\textbf{R}_i - \textbf{R}_j|^5}\right)\, ,\nonumber
\end{align}
where $\textbf{P}_i$, $\textbf{R}_i$ and $r_i$ are the first quantized momenta, position and transverse radial coordinates of the particles in the trap. $\omega_\bot$ is the transverse harmonic confinement frequency $m$ is the mass and $d$ the dipole moment of the particles.

A simple way to analyze the Hamiltonian (\ref{Full_Hamiltonian}) is through a classical mean field theory. This approach consists in minimizing the potential energy while neglecting the kinetic terms that induce fluctuation in the positions. A zigzag transition is captured by a Landau-like expansion of the configuration energy in powers of the staggered distortion $r$ of the particles.\cite{Fishman2008}
For transverse confinement frequency $\w_\perp$ below the critical frequency
$\w_c=\sqrt{ {279 \zeta(5) d^2 \rho_0 ^5 / 8 m}}$,
the classical energy is minimized by a non vanishing distortion $r=\kappa\,\rho_0 ^{-1} \sqrt{1-(\omega_\bot / \omega_c)^2}$, where $\kappa =\sqrt{186 \zeta(5)/3175 \zeta(7)}\approx 0.245$, $\zeta(n)$ is the Riemann zeta function, and $\rho_0$ is the particle density. In reality, this criterion only marks the characteristic frequency around which a local zigzag distortion begins to develop. Quantum fluctuations  allow the zigzag to twist and turn and may ultimately destroy the long range order.

To describe the soft fluctuations of the zigzag we assume that each particle is distorted off the axis of the trap by the fixed classical value $r$, but is free to rotate around the axis by the angle $\f$. All our expansions below are valid when the distortion $r$ is small compared to the inter-particle distance, that is $\tlr\equiv r\rho_0$ should be taken to be a small dimensionless parameter. In section (\ref{sec:mf}) we will see that in the relevant range of parameters for dipolar molecules this is always satisfied near the zigzag quantum phase transition.

The effective Hamiltonian which describes the linear motion of particles along the trap axis, rotations around the axis at a fixed radius $r$, and dipolar interactions between particles is given by
\begin{align}
H
&=\sum_i \bigg[ {p^2 _i \over 2 m}+\sum_{j > i} {d^2\over |x_i-x_j|^3}
+{L^2 _i \over 2 I}+  \label{First_quantized_Hamiltonian}\\&\sum_{j > i}{J_{ij}}\bigg(\cos(\f_i - \f_{j})+{1\over2} \cos(\f_i+\f_j)-{\nu\over 2}\cos2\f_i \bigg)\bigg] . \nonumber
\end{align}
Here $p_i$ and $x_i$ are the linear momenta and position operators and $L_i$ and $\f_i$ are the angular momenta (along the axis) and its conjugate angle operator of the $i$'th particle. The first two terms in the Hamiltonian (\ref{First_quantized_Hamiltonian}) describe the particle motion and interactions along the trap axis, which give rise to density fluctuations.
The rest of the Hamiltonian describes the fluctuations that drive the Ising transition. Apart from the last term, the angular part of the Hamiltonian looks like a quantum $xy$ model. The dipolar interaction, through the cosine coupling favors a staggered arrangement of the molecules off the axis, while the angular kinetic energy  delocalizes the angle and thereby acts to disorder the zigzag. The last term  breaks the $U(1)$ angle symmetry down to $\mathds{Z}_2$ by favoring the up and down ($0$ and $\pi$) distortions of the particles. If the trap potential has cylindrical symmetry then the preferred axis is set only by the external electric field which polarizes the dipoles. In this case we have $\nu=1$. It is possible to change $\nu$ by tuning the ratio between the transverse trap frequencies parallel and perpendicular to the electric field.

The other coupling constants appearing in the effective Hamiltonian (\ref{First_quantized_Hamiltonian}) are calculated directly from the full Hamiltonian (\ref{Full_Hamiltonian}). They  are given by $J_{ij} = J / (\rho_0 |x_{i}-x_{j}|) ^5$ where $J=6 d^2 r^2 \rho_0 ^5$ and $I=m r^2$. Note that the strength of the coupling $J_{ij}$ depends on the particles positions such that it produces a coupling between the angular and longitudinal degrees of freedom.
At this point it is convenient to express $J$ and $I$ in terms of two important dimensionless parameters that can be independently tuned in the system: (i) $\tilde{r}\equiv r\rho_0$ is the ratio of the classical distortion radius to the average inter particle distance and (ii) The dimensionless dipolar interaction strength $R_s=d^2\rho_0 m/\hbar^2$ is the ratio of the typical dipolar energy $\e_d=d^2 \rho_0^3$ to the typical kinetic energy $\e_0=\hbar^2\rho_0^2/m$. $R_s$ can be varied for example by changing the dipole moment $d$ using an external electric field, while $\tlr$ is tuned by varying the transverse trap frequency $\w_\perp$. We can express the Hamiltonian parameters using these dimensionless numbers as:
$J=6R_s \tilde{r}^2 \e_0$ and $I=\hbar^2\tlr^2/\e_0$.

\section{Estimation of the transition point}\label{sec:mf}
Within a classical analysis\cite{Fishman2008} the zigzag transition takes  place at a critical value of the transverse trap frequency given by $\w_c=6.014 \sqrt{R_s} \e_0/\hbar$. Quantum fluctuations driven by the kinetic energy in the Hamiltonian (\ref{Full_Hamiltonian}) or (\ref{First_quantized_Hamiltonian}) will lead to a transition at a smaller value of the transverse trap frequency. With increasing strength of interaction $R_s$ the relative importance of the kinetic energy decreases and we expect the critical $\w_\perp$ to approach the classical value.

We shall now formulate a quantum mean field theory of the effective Hamiltonian (\ref{First_quantized_Hamiltonian}) in order to estimate the value of the parameters at the transition point and to asses the importance of quantum fluctuations.
For this purpose we freeze the axial degrees of freedom and focus on the angular part of the Hamiltonian (\ref{First_quantized_Hamiltonian}). In addition since $J_{ij}$ decays rapidly with distance we consider only nearest neighbor interactions.

The mean field approximation consists of decoupling the interaction term in the Hamiltonian (\ref{First_quantized_Hamiltonian}) to get the local Hamiltonian
\begin{align}
H_{MF}=\sum_i \bigg[{L^2 _i \over 2 I}-{J}\bigg(3\,\s\,\cos\f_i  +{\nu\over 2}\cos2\f_i \bigg)\bigg]\,.\label{MF_Hamiltonian_SM}
\end{align}
This is supplemented by the self consistency condition
\begin{align}
\s = \langle \cos \f_i \rangle_{MF}\, .
\end{align}
Note that the Hamiltonian (\ref{MF_Hamiltonian_SM}) is fully quantum mechanical in the sense that the non-commuting variables $\f_i$ and $L_i$ are both present in it, in the interaction and kinetic terms respectively. Furthermore the transition is a result of competition between these two terms, and not driven by interaction alone as in the classical zigzag instability. On the other hand this local mean field scheme neglects the spatial structure of the fluctuations and hence cannot capture the universal critical behavior close to the transition. Below we estimate the fluctuation region in which the mean field theory fails.

Within the mean field approximation, the system develops a non zero order parameter $\s$ at the critical value of the couplings $IJ =\a(\nu)$. For example we find $\a(1)=0.31$, $\a(0.5)=0.33$ and $\a(2)=0.28$. While larger values of the eccentricity can be achieved in experiment, our quantum rotor representation  is not suited to describe them quantitatively.

\begin{figure}
\centering
\includegraphics[width=8.6cm,height=5.75cm]{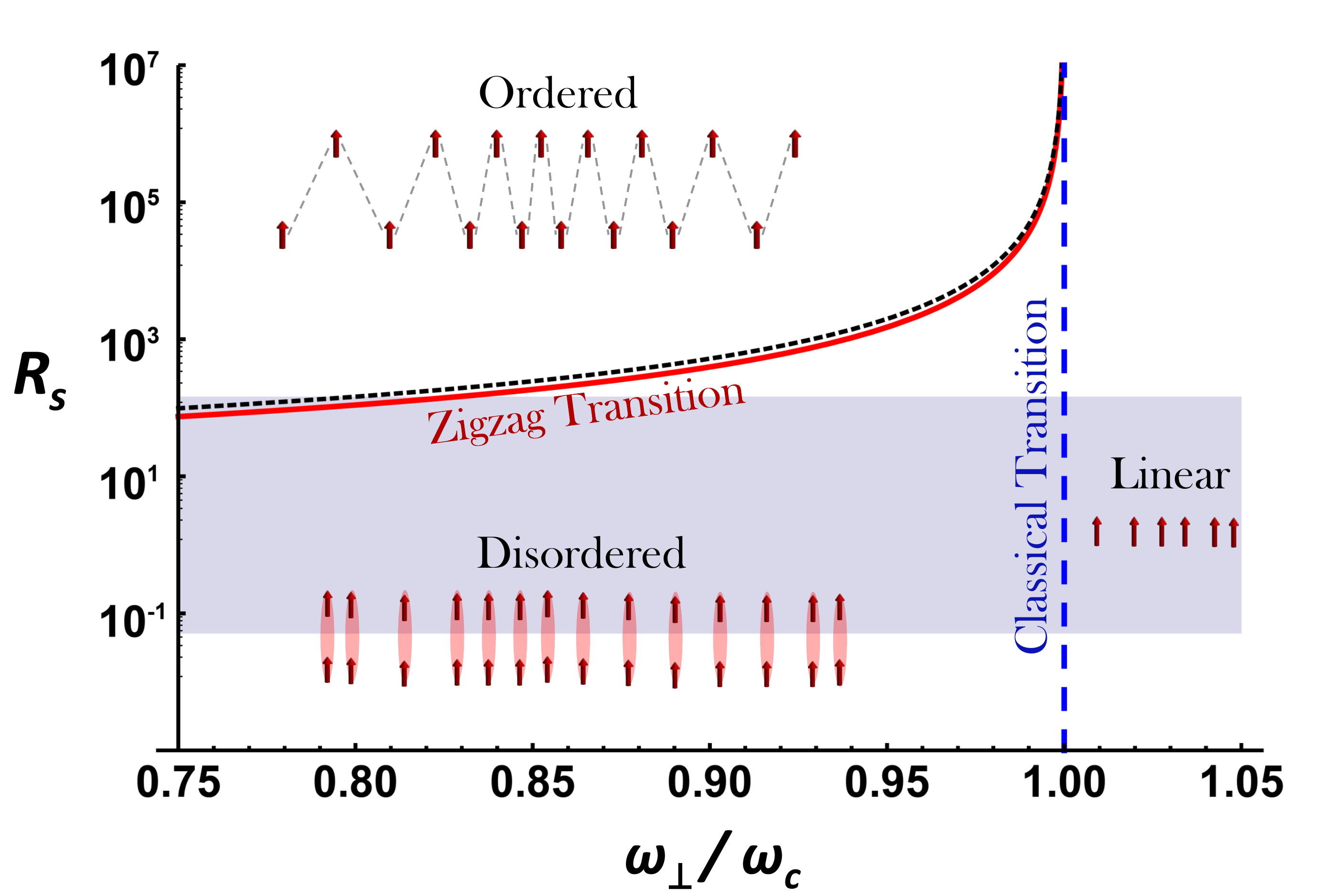}
\caption{ Phase diagram of polar molecules in an elongated trap in the space of dimensionless interaction constant, $R_s = d^2 m \rho_0$ and transverse confinement $\W_\bot / \W_c$, $\W_c$ is the critical transverse frequency in the classical limit. The solid curve marks the zigzag transition (\ref{Crit_line}) computed within the local mean field approximation for $\nu=1$ i.e. for isotropic transverse confinement. The dashed line corresponds to the Ginzburg criterion below which the we expect the mean field approximation to break down. Realistic values of $R_s$ range between $0.05$ and $130$ as marked by the grey band. Note that the quantum disordered zigzag and the linear chain are not sharply distinct. There is a smooth crossover between the two regimes. } \label{MF Phase Digaram}
\end{figure}

Using the relations $IJ=6R_s \tlr^4$ and $\tlr=\kappa \sqrt{1-(\w_\perp/\w_c)^2}$ we infer a phase boundary
\be
R_s^{\text{crit}}=46\,\a(\nu)\left[{1\over 1-(\w_\perp/\w_c)^2}\right]^2.\label{Crit_line}
\ee
The phase diagram is presented in Fig.\ref{MF Phase Digaram} for the case of a symmetric trap ($\nu=1$). We see that the transition point indeed approaches the classical value of $\w_c$ in the strong interaction limit.\cite{Astrakharchik2009}
Reasonable values for the dimensionless interaction $R_s$ expected in real systems can be estimated from the typical dipolar moments ranging from $d\simeq0.05 \,\mrm{D}$ for magnetic moments through $d\simeq0.5\, \mrm{D}$ for polarized molecules and up to $d\simeq10\,\mrm{D}$ for atoms weakly dressed with Rydberg excitations. Assuming densities of $\rho_0 \sim 10^{4} \, \mrm{cm}^{-1} $, we get $R_s$ between $0.05$ and $130$. From the phase diagram (Fig. \ref{MF Phase Digaram}) it is clear that in this regime quantum fluctuations shift the transition to $\w_{\perp}^{\text QC}$ far below the classical value $\w_c$. This implies a huge effect of quantum fluctuations on the transition in the relevant parameter regime.

We asses the accuracy of the quantum mean field approximation used to compute the transition point by applying the appropriate Ginzburg criterion $\av{\d\s^2}/\av{\s}^2\sim 1$.
The shift of the mean field transition line from the line on which the Ginzburg criterion is satisfied (black short-dashed line in Fig. \ref{MF Phase Digaram}) estimates the error in determination of the critical point. In terms of the dimensionless coupling $\a=IJ$ the Ginzburg criterion is satisfied at $\a\approx 0.4$ (for $\nu=1$) compared to the mean field transition found at $\a=0.31$.

%However, these values are placed in the disordered phase below the transition line, which can only be reached if $R_s$ is increased in the experiments.

\section{Long Wavelength Theory}\label{sec:ft}
To capture the universal properties of the zigzag transition and the effects of the low energy longitudinal fluctuations, we derive an effective long wavelength description of the Hamiltonian (\ref{First_quantized_Hamiltonian}).
The longitudinal modes are treated in a standard way. We replace the particle position $x_j$ with the smooth displacement field $\phi_\rho(x_j)$ and the particle momenta by the conjugate field ${1\over\pi}\partial_x\theta_\rho(x_j)$. These replacements allow to take the continuum limit of the first line in (\ref{First_quantized_Hamiltonian}) to obtain
\begin{equation}
\cH_\rho={\hbar\, u_\rho\over 2\pi}\int dx\bigg[ K_\rho (\partial_x \theta_\rho)^2+{1 \over K_\rho}(\partial_x \phi_\rho)^2\bigg]\, ,
\label{H_rho}
\end{equation}
where $K_\rho = \pi/ \sqrt{12 R_s}$ and $ u_\rho = \sqrt{12 R_s} (\hbar\rho_0/ m)$. From here on we rescale the coordinate $x$ by the inter-particle distance $\rho_0^{-1}$ and rescale energies by the typical kinetic energy $\e_0=\hbar^2\rho_0^2/m$. After this transformation the velocity becomes dimensionless $u_{\rho}=\sqrt{12 R_s}$.

The long wavelength limit of the angular Hamiltonian should be taken with more care. Because the angle $\f_i$ has a staggered arrangement it cannot be directly replaced by a continuum field $\f(x_i)$. Instead we must make a staggered transformation
\begin{equation}
\phi_{\sigma,i}=(\f_i + \pi N_i)\mod 2\pi \, ,\label{Non_Local_transformation}
\end{equation}
where $N_i=\sum_{j}\Theta(x_i-x_j)$ counts the number of particles to the left of the $i$'th particle. Thus the new variable is slowly varying in space, with the price of being a non local operator.
% and $\Theta(x)$ is the step function.
Now we take the continuum limit through the substitution $\phi_{\sigma,j}\rightarrow\phi_{\sigma}(x_j)$ and $L_j \rightarrow {1\over\pi\rho_0}\partial_x \theta_\sigma(x_j)$. In addition, noting that $N_i\to \pi\rho_0 x-\phi_\rho(x)$ we can express the original local Ising distortion as a composite of the two slowly varying fields $\hat{\s}(x)=r\cos(\f(x))= r\cos(\phi_\s(x)+\phi_\rho(x)-\pi\rho_0 x)$.

The long wavelength Hamiltonian in the angular sector, including only the most relevant terms is given by
\begin{align}
\cH_\sigma={u_\sigma\over 2\pi}\int & dx\left[  K_\sigma (\partial_x \theta_\sigma)^2+{1 \over K_\sigma}(\partial_x \phi_\sigma)^2\right]\label{H_sigma}\\
+\int & dx (-g_1\cos2\phi_\sigma+g_2\cos2\t_\sigma)\,, \nonumber\\
\cH_c = -\lambda {2\rho_0 \over \pi^2}& \int  \, dx \, \partial_x \phi_\rho \left(1-\cos 2\phi_\sigma \right) \, .\label{Coupling_Term}
\end{align}
Here again we have rescaled the coordinate $x\to x\rho_0$ and energies $\e\to \e/ \e_0$ in order to work with dimensionless coefficients. The term $g_1$ originates directly from the last term in (\ref{First_quantized_Hamiltonian}), from which its (bare) value can be estimated to be $g_1 =  J=6R_s \tlr^2$. The term $g_2$ accounts for the compactness of the original angle variable. The value of the coefficient is the fugacity of kink-anti-kink pairs in $\phi_\s$. It is estimated from the action associated with tunneling between the two potential minima
$\phi_\sigma=0$ and $\pi$ through the barrier $J \cos 2\theta_\sigma$.\cite{Matveev2004} Within the WKB approximation we obtain
$g_2 \approx I^{-1} \,\mrm{e}^{-8 \sqrt{I J}}=\tlr^{-2}\exp(-8\sqrt{R_s \tlr^4})$. In addition we extract from  (\ref{First_quantized_Hamiltonian}) the bare value of the Luttinger parameter
 $K_\sigma ^{-1}=\pi\sqrt{IJ} = { \pi\sqrt{6 \tlr^4 R_s}}$ and the velocity $u_\sigma = u_\rho / \sqrt{2}=\sqrt{6 R_s}$.

Finally the term (\ref{Coupling_Term}) describes the coupling between the longitudinal phonons and the zigzag degrees of freedom. It is generated because the energy gain associated with a zigzag arrangement, proportional to $J_{ij}$, is modulated with the distance $x_i-x_j$ between neighboring atoms along the chain. The field $\partial_x\phi$ which couples to the zigzag energy in (\ref{Coupling_Term}) is simply the long wavelength limit of $x_i-x_j$. The value of the coupling constant is again extracted directly from (\ref{First_quantized_Hamiltonian}), $\lambda = 15\pi R_s \tlr^2$.
Other coupling terms between the two sectors have been omitted since they turn out to have lower scaling dimension at the critical point which describes the zigzag transition.

We note that our derivation of the field theory was carried out in the strongly interacting limit $R_s\gg 1$, where the particles almost form a Wigner crystal. In this case the low energy Hamiltonian is essentially independent of the statistics (Bosonic or fermionic) of the constituent particles. As we shall see later, the differences due to particle statistics arise when we try to probe the system by inserting or extracting single particles to it.

Having derived the effective field theory we are now in position to discuss the resulting quantum phases and phase transitions. Let us start with the phases.
The Hamiltonian $\cH_\s$ of the angular degrees of freedom describes a competition between two cosine terms in a regime where they are both relevant with respect to the quadratic part of the Hamiltonian. In the zigzag ordered phase the term $g_1$ is dominant. The field $\phi_\sigma$ is then pinned to one of the two minima at $0$ or $\pi$, and the Ising-like order parameter $\langle \cos\phi_\sigma(x) \rangle$ takes a non vanishing value $+1$ or $-1$. On the other hand when $g_2$ becomes dominant the dual variable $\theta_\s$ is pinned, leading to proliferation of phase slips in $\phi_s$ and disordering of the Ising field.

The representation of the order parameter in terms of the Bosonic field $\phi_\s$ conceals the fact that the order in the zigzag phase is non-local. It consists of an infinite string operator when written in terms of the actual particle distortion $\hat{\s}(x)=r\cos\f(x)$
\begin{equation}
\langle  r\cos\phi_\sigma(x) \rangle=\langle \mathrm{e}^{\mathrm{i} \pi \int_{-\infty} ^x \rho( x')\, dx'} \hat{\s}(x) \rangle\, ,\label{String_Order_Parameter}
\end{equation}
where $\rho(x)$ is the particle density.
%The integral in the exponent counts the number of particles to the left of the point $x$.
The string order parameter captures the fact that the phase of the zigzag, namely an up or down distortion, is alternating in the particle frame but is not fixed to a position in the lab frame. The direct correlation function of the transverse distortions $\hat{\s}(x)=r\cos\f(x)$ decays as a power-law along the chain due to disordering of the positional order:
\be
\av{\hat{\s}(x)\hat{\s}(0)}
\to r^2\cos(\pi\rho_0 x)\av{\cos\phi_\s}^2 |x|^{-K_\rho}
\ee

Let us move on to discuss the critical point separating the zigzag from the disordered phase. If we ignore for the moment the coupling $\lambda$ between the zigzag degrees of freedom and the longitudinal phonons, then by symmetry considerations the effective Hamiltonian (\ref{H_sigma}) is expected to describe an Ising critical point.
When $K_\sigma=1$ the problem becomes particularly simple, since we can immediately express $\cH_\s$ as a model of two independent Majorana Fermions or alternatively a single Dirac Fermion:\cite{Shelton1996,Lecheminant2002}\cite{}
\begin{equation}
\cH_\sigma = \int \sum_{\eta=\pm} dx\,\xi^\eta  \left[-\mathrm{i}\, {\tilde{u}_\s\over 2} \,\partial_x\,\tau ^z + \Delta_\eta \,\tau^y \right]\xi^\eta \, .\label{Fermion_Hamiltonian}
\end{equation}
Here $\tu_\s=u_\s $, $\xi^\eta  = (\,\xi_R ^\eta \,,\,\xi_L ^\eta\,)^T$ is a Majorana spinor in the chiral basis, and $\tau^{y,z}$ are Pauli matrices that act in this basis. The masses of the two Majorana modes are given by $\D_\pm=\pi(g_2\pm g_1)$. The transition at $g_1=g_2$ is then described by a single massless Majorana mode $\xi^-$, while the other gapped mode can be ignored. This is precisely the critical theory of the transverse field Ising model.\cite{Zuber1977}
The relation between the low energy Majorana mode and the Bosonic fields in the low energy limit is
\begin{eqnarray}
\xi_{R} ^-&=&{  -\sqrt{\rho_0 / \pi}}:\sin{(\theta_\sigma-\phi_\sigma)}: \label{Fermionization_Dictonary_R}\\
\xi_{L} ^-&=&{  \sqrt{\rho_0 / \pi}}:\cos{(\theta_\sigma+\phi_\sigma)}: \label{Fermionization_Dictonary_L}\,
\end{eqnarray}

From the mean field analysis presented in section (\ref{sec:mf}) we expect the transition to occur at a bare value of $K_\s$ close to, but not necessarily precisely one. A deviation from $K_\s=1$ (see appendix \ref{Fermionization_of_DSG_appendix}) gives rise to an interaction that couples the two sectors of Majorana Fermions
\begin{align}
\cH_\sigma ^{(I)} = V \int dx  \,\xi_R ^- \xi_L ^+ \xi_R ^+ \xi_L ^-\, ,\label{Hint}
\end{align}
where $V=2\pi u_\s \left(K_\s ^{-1}-K_\s\right)$. In addition there is an increase in the velocity of the Majorana modes, which now becomes $\tu_\s={u_\s\over 2}\left(K_\s^{-1}+K_\s\right)$.

Because the interaction couples the critical modes to a gapped excitation it is perturbatively irrelevant at the critical point. In other words the double sine-Gordon model (\ref{H_sigma}) flows under renormalization to the self dual (Ising) point $K_\s=1$. The residual effect of the interaction is to shift the transition away from the point $g_1=g_2$. This shift can be estimated using a mean field decoupling of $\xi^-_R\xi^-_L$ from $\xi^+_L\xi^+_R$ in (\ref{Hint}).

So far we have ignored the coupling $\lambda$ between the Ising degrees of freedom and the longitudinal phonons. The effect of this coupling, which a priori breaks the conformal invariance and Lorentz-symmetry of the Ising critical point, was addressed in Ref. [\onlinecite{Sitte2009}] using a perturbative renormalization group analysis in $\lambda$. To one loop order, $\lambda$ was found to be irrelevant if the sound velocity $u_\rho$ is initially larger than the velocity of the Majorana mode $\tu_\s$. The slow flow toward $\lambda=0$ has a peculiar effect on the fixed point. The ratio between the two velocities $\tu_\s/ u_\rho$ approaches unity, thereby restoring Lorentz symmetry at the fixed point. However, at the same time the velocities themselves flow to zero. In the opposite regime, $\tu_\s> u_\rho$ the coupling $\lambda$ is relevant. The nature of the transition in this case is not well understood and furthermore corrections beyond one loop can change the picture. We would like to point out however that a perturbative expansion in $\lambda$ is justified for the bare values of the parameters dictated by our system. The small parameter of the expansion is $\lambda\sqrt{K_\rho/(\tu_\s u_\rho)}$, which in our case scales as $1/\sqrt{R_s} \ll 1$.

An experiment can in principle explore both of the above regimes by tuning the ratio  $\tu_\s /u_\rho$. We have shown in section \ref{sec:mf} that changing the eccentricity of the transverse confinement leads to change of the critical value of $R_s \tlr^2$ and therefore also of the bare Luttinger parameter $K_\s$ near the transition point. For $K_\s=1$ the ratio $\tu_\s/u_\rho=1/\sqrt{2}<1$, while by changing (bare) $K_\s$ the ratio can be tuned above $1$.

\begin{figure*}
\centering
\includegraphics[width=18cm,height=3.3cm]{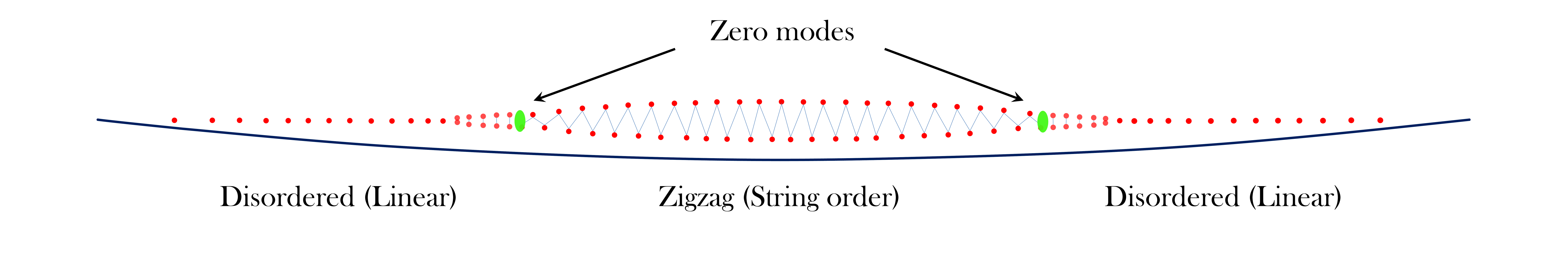}
\caption{States of dipolar particles along an elongated trap with soft harmonic confinement. The inhomogeneous density along the trap acts as a tuning parameter of the zigzag transition. Here the particles in the dense middle section are in the zigzag phase while the wings of the cloud are in the disordered phase. Majorana zero modes are exponentially localized near the interfaces between the two phases. }\label{Realistic:fig}
\end{figure*}

Finally, it is interesting to note the difference between our model of the zigzag transition and the problem of a two leg Fermion ladder discussed in Ref. [\onlinecite{Starykh2000}]. As we have mentioned above, such a Fermion model should not be essentially different from ours in the strong interaction limit. However Ref. [\onlinecite{Starykh2000}] assumes the weak coupling perspective, taking the Fermi energy to be much larger than both the interaction and the splitting between the two sub-bands. This leads after Bosonization (using our conventions) to a double sine-Gordon model similar to (\ref{H_sigma}) in the ``spin'' (antisymmetric) sector, but with the term $g_2 \cos2\t$ replaced by a $\cos 4\t$ term, which corresponds to Cooper pairing. Accordingly the phase competing with the zigzag in the weak coupling limit is a Superconductor (i.e. quasi long range order in the pairing field), while in our case the competing phase is unpaired. The critical theory which describes the transition into the paired phase is different from the transition to the the unpaired state found in our case. Note in particular that the self dual point of the double sine-Gordon model with the pairing term is at $K_\s=2$ rather than $K_\s=1$ in our case.

\section{Zero energy edge states}\label{sec:edge}

The non-local order has interesting consequences for the physics of the edge.
Let us consider a physical model of the edge as an interface between the string-ordered zigzag phase and the disordered phase, created by a spatially dependent gap parameter $\D(x)$ that changes sign on the interface. This is in fact a realistic model for dipolar particles in a harmonic trap (see FIG. \ref{Realistic:fig}). The density which slowly decreases away from the center along the trap is a tuning parameter for the zigzag instability. Therefore, the dense middle section may be in the zigzag phase, the wings of the cloud are in the disordered phase, and there is necessarily an interface between the two regions.

The Hamiltonian (\ref{Fermion_Hamiltonian}) is formally identical to a one dimensional Superconductor of spinless Fermions. This model has a zero-energy Majorana mode localized at the interface, where the gap function changes sign.\cite{Kitaev2001} The zero-mode has a particularly simple structure if $\D$ changes abruptly across the interface from $\D_0$ to $-\D_0$ (see appendix \ref{app:edge}). It is then given by the Majorana operator
\begin{equation}
\g_0=\sqrt{1\over 2\,l}\,\int dx\, \mrm{e}^{-{|x|\over 2l}}\left(\xi_L(x)+\xi_R(x)\right)\,.
\end{equation}
where $l =\tu_\s /4 \Delta_0 $ is the correlation length.
Bulk states of (\ref{Fermion_Hamiltonian}) are above the gap $\Delta_0$.

Here we should remark on the difference between the Majorana edge modes of the zigzag phase and those found in two other one-dimensional systems. Consider first the transverse field Ising model. The representation of this  model in terms of Jordan-Wigner Fermions shows  Majorana zero modes localized at the two edges. But when mapped back to the original Ising degrees of freedom, these two modes simply span the Hilbert space of the two broken symmetry states in the bulk. In our case, because there is no broken symmetry in the bulk the zero modes are truly localized at the edge, as we will show explicitly below.

There is also an important difference between the Majorana edge modes discussed here and those of a spinless Fermion model in which the Fermions are the local degrees of freedom. In our case, the Majorana modes appear only in the spin-sector, where they describe a double degeneracy of the spectrum for a given charge parity of the zigzag. For even parity, the two states can be understood as either a zigzag that starts with an up distortion and ends down, or the opposite configuration. For odd parity the two states correspond to a zigzag with an up or down distortion at both ends. However in a system with an interface, where the parity is not well defined, all four states should exist and may be combined into completely local excitations. This is in contrast to the non-local nature of the edge modes in the Fermionic wire.\cite{Kitaev2001}

We shall look for physical signatures of the edge states in the local tunneling density of states (DOS).\cite{Starykh2000} Consider a probe that can tunnel particles into or out of the system at a desired point $x$ and energy $\w$. Such local probes have been proposed\cite{Kollath2007b}, and are now becoming a reality in cold atomic systems.\cite{Bakr2010,Sherson2010}
We also assume that the position of the tunneling tip is asymmetric with respect to the tube axis so that it can insert a particle either displaced up or ($s=\ua$) down ($s=\da$) in the tube. The tunneling rate is proportional to the local spectral function
\begin{align}
A_s(\omega,x) =  \mrm{Im}\left[{\mrm{i}\over \pi}\,\int dt\, \mrm{e}^{\mrm{i} \omega t} \Theta(t)\langle \left[\, \Psi_s(t,x) \,,\,\Psi^\dag_s (0,x)\, \right]_\pm\rangle\right].\label{Local_DOS}
\end{align}
where the $\pm$ subscript denotes anti-commutator and commutator for Fermions and Bosons respectively. To find the low energy behavior of the spectral function we seek the representation of the particle field operators in terms of the Bosonic density modes $\phi_\rho$ and $\theta_\rho$ and the low energy Majorana Fermion $\xi$. This will depend in a crucial way on the species,
i.e. Bosonic or Fermionic, of the dipolar particles. In what follows we separately treat each of these cases.

{\em Bosonic particles--} Let us start from the standard Bosonization identity for Bosonic particles in two chains labeled by $s=\ua ~(+),\da ~(-)$
\begin{align}
\Psi\yd_{B,s} \simeq \sqrt{\rho_0} \mrm{e}^{-\mrm{i}\,\theta_s}\sum_{m=-\infty} ^\infty \mrm{e}^{-\mrm{i}2m(\phi_s+\pi\rho_0 x )}\,.\label{Bosonic_Operator}
\end{align}
We now canonically transform to a representation in terms of symmetric (``charge'') and anti-symmetric (``spin'') fields through $ \theta_s = \theta_\rho + s\, \theta_\sigma$ and $\phi_s = (\phi_\rho + s\, \phi_\sigma)/2$. Keeping the most relevant terms $m=0,\pm 1$ and using the Bosonized form (\ref{Fermionization_Dictonary_R},\ref{Fermionization_Dictonary_L}) of the Majorana operators we obtain
\begin{align}
{\Psi\yd_{B,s}}&\simeq  \b_0 \,\sqrt{\rho_0}\, \mrm{e}^{-\mrm{i}\,\theta_\rho }\mrm{e}^{-\mrm{i}\, s\,\theta_\sigma}  \label{Bosonic_Operator_3} \\
 &+\b_1\,\mrm{e}^{-\mrm{i}\,\theta_\rho }\left(\mrm{e}^{\mrm{i}(\phi_\rho+2\pi\rho_0 x) } \,\xi_L -\mrm{i}\, s \,\,\mrm{e}^{-\mrm{i}(\phi_\rho +2\pi\rho_0 x)}\, \xi_R  \right)\, .\nonumber
\end{align}
where $\b_0$ and $\b_1$ are non-universal coefficients.
The first row in the operator (\ref{Bosonic_Operator_3}) corresponds to the long-wavelength components of the Boson. The second row describes the components with wavelength comparable to the inter-particle spacing. In both terms insertion of a Boson must cause a disruption in the zigzag order. In the first case the operator $\mrm{e}^{\pm \mrm{i}\,\t_\s}$ creates a kink in the Ising order-parameter, while in the second case the disruption is affected by the Majorana operator.

Since we are interested in the low energy contributions near the interface we can further simplify the Bose operator. First the factor $\mrm{e}^{{\pm \mrm{i}\,\t_\s}}$ can be replaced by its expectation value,
which decays exponentially with distance from the interface in the ordered side (recall that $\av{\mrm{e}^{\mrm{i}\,\t_\s}}$ is the ``dual Ising order-paramete''). Second, the Majorana operators $\xi_{L,R}$ can be replaced by the contribution to them from the zero energy edge mode, $\xi_{L,R}\sim   \mrm{e}^{-{2\D_0\over  v}|x|}\gamma_0$. From here we can easily compute the local spectral function at low energies (see appendix.\ref{Tunneling_DOS_appendix})
\be
A(E,x)\approx \left({{B_0}\over 1+\mrm{e}^{\,{x/ l}}} + B_1\, \mrm{e}^{-{|x|/ l}} |E|^{K_\rho}\right)|E|^{{1\over 4K_\rho}-1}\,,
\ee
where $B_{0,1}$ are non-universal constants. The leading (first) term originates from the long wavelength contribution to the Bose operator. Only the
 sub-leading term $B_1$ stems from the Majorana  zero mode.

{\em Fermionic particles --} Unlike with Bosonic particles, we cannot access the zigzag phase of Fermions starting from weakly coupled Luttinger liquids. A better starting point to obtain the low-energy limit of the Fermion field is a two band model, including the ground state and first transverse excitation in the tube.\cite{Meyer2007,Meyer2009,Sitte2009} Details of the formulation are left to appendix \ref{Conneting_to_low_energy_Fermions}, however the end result is simple. A Fermion inserted into the lower band is described by
\be
\psi_0 ^\dag \simeq  \a_0\, \sqrt{\rho_0}\sum_{r=R,L}\mrm{e}^{-\mrm{i} \, r k_F x}  \mrm{e}^{\mrm{i} \, (r \phi_\rho -\theta_\rho)}\mrm{e}^{{-\mrm{i} \,r\, \theta_\sigma}}
\ee
This is directly analogous to the first term in (\ref{Bosonic_Operator_3}), except that the Fermion is not inserted at zero momentum but rather at the Fermi points of the Luttinger liquid that describes the breathing modes of the zigzag. Insertion of a Fermion must be accompanied by creation of a kink in the zigzag order, which is implemented by the factor $\mrm{e}^{{\pm \mrm{i}\,\t_\s}}$. Because on both sides of the transition there is a gap to a second band, the Fermion can also be inserted at zero momentum above the gap. This ``second band'' Fermion is directly proportional to the Majorana modes (see appendix \ref{Conneting_to_low_energy_Fermions})
\be
\psi\yd_1\simeq \a_1\,{\mrm{e}^{\,\mrm{i} {\pi\over 4}}\over \sqrt{2}}\, \left( \xi_R  - \mrm{i}\, \xi_L  \right)\,\mrm{e}^{-\mrm{i}\,\t_\rho}\,.
\ee
A Fermion distorted up or down in the tube is now described by a superposition of the symmetric and antisymmetric transverse states
so that $\Psi_{F,s} = (\psi_0+ s\,\psi_1)/\sqrt{2}$.

With the Fermion operators at hand we are now in position to compute the local spectral function related to the tunneling rate of up or down distorted Fermions into a point $x$. The result is again a power-law
\be
 A(E,x)\approx \left({{A_0 }\over 1+\mrm{e}^{\,{x/ l}}}|E|^{K_\rho} + A_1\, \mrm{e}^{-{|x|/ l}}\right)|E|^{{1\over 4K_\rho}-1}\,. \label{Fermi_DOS}
 \ee
As in the Bosonic case, the first term describes the ``penetration'' of the disorder-parameter into the ordered phase near the interface.
The second term is the contribution of the Majorana mode. Note however that contrary to the Bosonic case, here the Majorana mode gives the dominant contribution to the low energy tunneling rate near the interface.
It is interesting to note that Eq. (\ref{Fermi_DOS}) shows the same leading behavior with $E$ as found in Ref. [\onlinecite{Starykh2000}] for tunneling into the edge of the zigzag phase in the weak coupling limit. This is in spite of the very different modeling and physical picture of the edge in the two scenarios. From our analysis it is clear that this behavior stems from a robust zero energy state with topological origin.

\section{Numerical results}\label{sec:dmrg}
\begin{figure}
\begin{center}
\includegraphics[width=9cm,height=6.3cm]{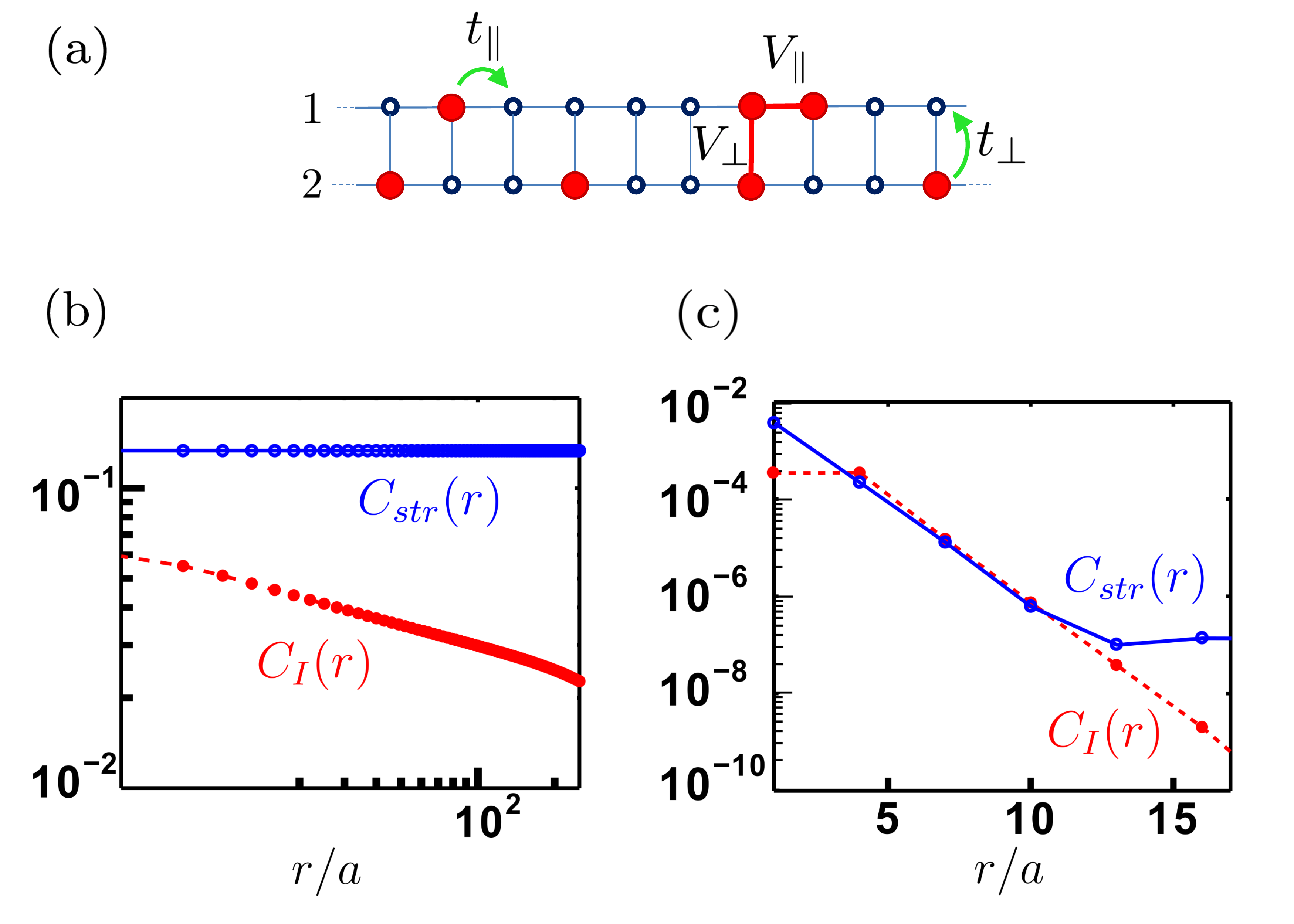}
\end{center}
\caption{(Color online) DMRG calculation of correlations in the two-leg ladder model (\ref{DMRG_Hamiltonian}). The two legs of the ladder illustrated in panel (a), correspond to the up and down distortion. $V_\perp$ is taken to be the largest energy scale to exclude double occupancy of a rung, and the zigzag transition is tuned by varying the ratio $t_\perp/V_\parallel$. (b) The component with spatial frequency $\pi\rho_0$ of the Ising (red dashed) and string (blue solid) correlation functions in the ordered phase on a Log-Log plot. The string correlations display true long range order while the Ising correlations decay as a power-law.  (c) The same correlation functions calculated in the disordered phase plotted on a Log-Linear plot, where both are seen to decay exponentially.}\label{fig.5}
\end{figure}

We now turn to a numerical calculation of a concrete microscopic model using DMRG\cite{White1992} and verify that the long range correlations  are indeed well described by the field theoretical analysis presented above. For this purpose we consider a lattice model which describes similar physics as the original Hamiltonian (\ref{First_quantized_Hamiltonian}). Specifically we take a model of  hard-core Bosons on a two leg ladder at low incommensurate filling (see Fig. \ref{fig.5}.(a)):
\begin{align}
&H_{Ladder} = \sum_{i=1} ^{L} \bigg[-t_\parallel \sum_{a=1,2}(b_{i,a} ^\dag b_{i+1,a} +h.c.)\label{DMRG_Hamiltonian}\\
&-t_\perp (b_{i,1} ^\dag b_{i,2}+h.c )+
\sum_{a=1,2}V_\parallel  n_{i,a} n_{i+1,a}  + V_\bot n_{i,1} n_{i,2} \bigg]\, ,\nonumber
\end{align}
where $b^\dagger_{i,a}$ create a (hard-core) Boson on site $i$ of leg $a=1,2$. The two legs of the ladder represent the preferred up and down distortion of a particle. The corresponding Ising variable is the relative density $\hat \s_i=n_{i,1}-n_{i,2}$. The repulsive interaction $V_\parallel$ between particles on the same leg favors a zigzag arrangement to minimize the interaction energy, just as the dipolar interaction does in the original system. At the same time the hopping on a rung $t_\perp$ counters that ordering tendency. The latter is directly related to the tunneling matrix element between the two favored states of the rotor $\phi_\s=0$ and $\pi$. Tuning $t_\perp$ to drive the phase transition is analogous to varying the transverse trap frequency in the original model. The hopping $t_\parallel$ along the ladder drives the longitudinal quantum fluctuations. At incommensurate filling, just as in the continuum, these fluctuations prevent crystalline order from forming. Finally we fix a large repulsion  $V_\perp$ ($\gg V_\parallel,t_\parallel, t_\perp$) to suppress double occupation of a rung.

To characterize the ground state we compute the staggered Ising correlation function and string correlation function, which are defined respectively as:
\begin{align}
&C_I({r_i-r_j})=\cos\left(\pi\rho_0 (r_i-r_j)\right)\av{\hat{\s}_i \hat{\s}_j} \nn\\
&C_{str}({r_i-r_j})=\av{\hat{\s}_ie^{\mrm{i}\,\pi\sum_{l=j} ^i n_l}\hat{\s}_j}\,.
\end{align}
Results of two calculations which differ only by the value of the tuning parameter $t_\perp$ are presented in Fig. \ref{fig.5}. The lattice size in both cases is
$L=256$ unit cells (rungs), the filling is $\rho_0 \approx 1/3$ per unit cell, and $V_\parallel=t_\parallel=1$. The transverse hopping is chosen to be $t_\perp=0.05$ in one calculation (panel (b) of the figure) and $t_\perp=0.8$ in the other (panel (c)), so that the system is in the zigzag ordered phase in the first and in the disordered phase in the second calculation. As anticipated, the Ising correlations decay like a power law of distance in the ordered phase while the string correlation saturates to a finite value. When the transverse tunneling is increased to $t_\perp=0.8$ both correlations decay exponentially with distance.

\section{Conclusion}\label{sec:summary}

The zigzag transition observed in chains of trapped ions\cite{Birkl1992} is a mechanical distortion that occurs when the ion chain is compressed or the transverse confinement decreased beyond a critical point. The transition is well described by a classical theory neglecting all quantum (as well as thermal) fluctuations.\cite{Fishman2008} In this paper we have shown how the analogous transition takes place in an elongated trap of ultracold dipolar particles, where quantum fluctuations are greatly enhanced compared to the ion system.

Quantum fluctuations are found to have a profound effect on the transition and on the nature of the ordered phase. In particular, the $\mathds{Z}_2$ symmetry of the system, which is broken in the classical zigzag phase, is restored by the quantum fluctuations of particle positions along the trap.  Instead, the quantum zigzag phase is characterized by a non-local, Ising string order parameter.
We also find zero energy edge states, localized at interfaces between the zigzag phase and the disordered (linear) phase, that are a direct consequence of the non local order in the bulk. It is notable that such non trivial correlations are found in the most natural setting for polar molecules with no need for special engineering of interactions.

We have calculated the experimental signatures of the edge states that are expected to be seen in tunneling experiments. It would also be interesting to observe the bulk string order directly using in situ detection of the particle positions.\cite{Endres2011} From the fundamental perspective, our analysis provides another example of a rather rare class of systems which exhibit topological order (or at least non-local string order) in a gapless phase. Other examples in this class include the Haldane phase of spin-$3/2$ chains and a Haldane liquid phase predicted to occur for two component dipolar Fermions in an optical lattice.\cite{Kestner2011} The zigzag phase is somewhat special in this class in that it occurs in presence of full (continuous) translational symmetry rather than in a lattice.

{\it Acknowledgments.} We thank P. Azaria, E. Berg, T. Giamarchi and A. Vishwanath for helpful discussions. E.A. Acknowledges support of the ISF and of the Louis and Ida Rich Career development chair. S.D.H  acknowledges support by the Swiss Society of Friends of
the Weizmann Institute of Science. EGDT is supported by the Adams Fellowship Program of the Israel Academy of Sciences and Humanities.

\appendix

\section{Re-Fermionization of the double sine-Gordon model}\label{Fermionization_of_DSG_appendix}
Here we review the re-Fermionization scheme for the double sine-Gordon model\cite{Shelton1996,Lecheminant2002,Bosonization_Books}(\ref{H_sigma}). %In addition we will discuss the out comes of re-Fermionization away from the point $K_\sigma = 1$.
Let us start by re-writing the Hamiltonian (\ref{H_sigma}) as follows:
\begin{align}
\cH_\sigma={u_\sigma\over 2\pi}{1+K_\sigma ^2 \over 2 K_\sigma}\int & dx\left[   (\partial_x \theta_\sigma)^2+(\partial_x \phi_\sigma)^2\right] \label{re_write_Ham_Sigma} \\
+{u_\sigma\over 2\pi}{1-K_\sigma ^2 \over 2 K_\sigma}\int & dx\left[- (\partial_x \theta_\sigma)^2+(\partial_x \phi_\sigma)^2\right]\nonumber \\
-\int & dx (g_1\cos2\phi_\sigma+g_2\cos2\t_\sigma)\, . \nonumber
\end{align}
Using the identities
\begin{eqnarray}
&&{  \sqrt{\rho_0 / \pi}}:\mrm{e}^{-\mrm{i}(\theta_\sigma-\phi_\sigma)}:\simeq \xi_{R} ^++\mrm{i}\,\xi_{R} ^-\label{Fermionization_Dictonary_R_Ap}\\
&&{  \sqrt{\rho_0 / \pi}}:\mrm{e}^{-\mrm{i}(\theta_\sigma+\phi_\sigma)}:\simeq\xi_{L} ^-+\mrm{i}\,\xi_{L} ^+\label{Fermionization_Dictonary_L_Ap}\, ,
\end{eqnarray}
we find that the re-Fermionized version of the first and third terms in this Hamiltonian (\ref{re_write_Ham_Sigma}) give the quadratic Majorana model
\begin{equation}
\cH_\sigma ^{(0)}= \int dx \sum_{ \eta=\pm}\xi ^\eta \left[-\mrm{i} \,{\tu_\s \over 2}\, \partial_x\tau ^z + \Delta_\eta \tau^y \right]\xi ^\eta\, ,\label{Fermion_Hamiltonian_supp}
\end{equation}
where $\xi^\pm = (\, \xi_R ^\pm\,,\,\xi_L ^\pm\,)^T$, $ \tu_\s =  {u_\sigma}(1+K_\sigma ^2 )/2 K_\sigma$ and $\Delta_\pm = \pi (g_2 \pm g_1)/\rho_0 $.
The re-Fermionized version of the second term in (\ref{re_write_Ham_Sigma}) has the form of an interaction
\begin{align}
\cH_\sigma ^{(I)} = V \int dx  \,\xi_R ^- \xi_L ^+ \xi_R ^+ \xi_L ^-\,.\label{Interaction_term}
\end{align}
where $V = {2 \pi u_\sigma}(1-K_\sigma ^2 )/ 2 K_\sigma$. As explained in section \ref{sec:ft} this interaction is irrelevant in the renormalization group sense and therefore not expected to alter the critical properties. It can however affect a shift of the critical point from the naive value $\D_-=0$. As explained in the text this shift can be estimated by decoupling the interaction term $V$ so that the renormalized values of $\D_+$ and $\D_-$ become mean field parameters to be determined self consistently. If $V$ is small compared to $\D_+$ the shift in $\D_-$ can be easily obtained perturbatively by substituting $\xi_R ^+ \xi_L^+$ in the interaction term by their expectation value in the unperturbed ground state
\begin{align}
-\mrm{i}\langle \xi_R ^+ \xi_L^+ \rangle = \,{\Delta_+ \over \rho_0 \tu_\s}\log \left[ {{\rho_0 \tu_\s\over \Delta_+}+\sqrt{\left({\rho_0 \tu_\s\over \Delta_+}\right)^2+1}}\right].
\end{align}
Then the renormalized value of $\D_-$ is $\tilde{\D}_-=\D_- -\mrm{i}\langle \xi_R ^+ \xi_L^+ \rangle V$. In the limit $\D_+\gg\rho_0 \tu_\s$ this approaches $\tilde{\D}_-=\D_- +V$.

\section{ The Majorana edge state}\label{app:edge}
In this appendix we solve the Bogoliubov-deGennes (BdG) equations in the presence of an edge and show that there is a single zero energy solution. For simplicity we take a sharp change in the mass term
\begin{equation}
\Delta_-(x) = \bigg\{
\begin{matrix}
\Delta_0 & x<0 & \Rightarrow & \mrm{disorderd}\\
-\Delta_0 & x>0& \Rightarrow & \mrm{orderd}
\end{matrix}\label{Perturbations_Spatial_Dependence}\,.
\end{equation}
When $K_\sigma = 1$ the critical point of the Ising model (\ref{H_sigma}) is at $\Delta_- =0$, therefore, equation (\ref{Perturbations_Spatial_Dependence}) defines a boundary between an ordered phase for $x>0$ and a disordered one for $x<0$. The BdG equations for this situation are given by
\begin{align}
\left[ -\mrm{i}\,{u_\s \over 2} \partial_x \tau^z +\Delta_-(x)\tau^y  \right]\,\chi_E(x) = E\, \chi_E(x) \label{BdG_eq}
\end{align}
where $\chi_E(x) = \left\{u_E(x),v_E(x)\right\}^{\mrm{T}}$ is the eigenstate of the equation (\ref{BdG_eq}) with energy $E\,$.  The physical solutions of (\ref{BdG_eq}) are at energies above the gap ($E>\Delta_0$) except for the single zero energy solution
\begin{equation}
\chi_0(x) = {\sqrt{2\Delta_0 \over  u_\s} \mrm{e}^{-{2\Delta_0 \over  u_\s}|x|}}
\begin{pmatrix}
1 \\
-1
\end{pmatrix} \, ,\label{Zero_Energy_State}
\end{equation}
The solution of the BdG equations immediately gives us the quasi-particle operators
\begin{align}
\gamma_E = \int \, dx \, \left[ u_E (x) \, \xi_R (x) - v_E(x) \, \xi_L (x)\right]\\
\gamma_E ^\dag = \int \, dx \, \left[ u_E ^* (x) \, \xi_R (x) - v_E ^*(x) \, \xi_L (x)\right]
\end{align}
and especially the localized Majorana operator
\begin{align}
\gamma_0  = \gamma_0 ^\dag={\sqrt{2\Delta_0 \over u_\sigma} \int dx\,\, \mrm{e}^{-{2\Delta_0 \over u_\sigma}|x|}} \,\left(\xi_R ^- + \xi_L ^-\right)\,.\label{Bogoliubov_Transformation_R}
\end{align}

\section{Low energy limit for Fermionic dipoles}\label{Conneting_to_low_energy_Fermions}
In case of Fermions it is simpler to derive the Bosonization identities starting from the weak coupling picture
considered in Refs. [\onlinecite{Meyer2007,Meyer2009,Sitte2009}]. Here we will review the derivation of the low energy theory in this case and draw analogies to the field theory derived above (\ref{H_sigma}) starting from strongly coupled dipolar particles. Guided by this low-energy theory we will give a phenomenological expression for the field operators.

In Ref. [\onlinecite{Sitte2009}] the two lowest sub-bands of the transverse confining potential are considered. The zigzag transition takes place when the chemical potential is placed at the bottom of the second sub-band. At this point the group velocity of the second sub-band is zero and the Bosonization procedure can not be carried out. However, we can still Bosonize the first sub-band.
The Hamiltonian of the two sub-bands can be written as follows
\begin{align}
H = {u_0 \over 2\pi}& \int dx \left[ K (\partial_x \theta_0) ^2 +{1\over K} (\partial_x \phi_0) ^2 \right]\label{Matveev_Hamiltonian} \\
 +&\int dx \,\psi_1 ^\dag \left( -{\partial_x ^2 \over 2m} + \d \right) \psi_1\nonumber \\
 +&\int dx \left[ - {g\over \pi} \partial_x \phi_0 \psi_1 ^\dag \psi_1 +{u_1\over 2} (\mrm{e}^{\mrm{i} \, 2 \theta_0} \psi_1 \partial_x \psi_1 +h.c.)\right]\nonumber
\end{align}
where $\d$ is the difference between the energy of the second sub-band and the chemical potential, $\d = \e_1 - \mu$.
The field operator $\psi_ 1$ belongs to the second sub-band and the first sub-band has been Bosonized
\begin{align}
\psi_0 = \sum_{r=R,L}\mrm{U}_r\lim_{\rho_0 \rightarrow \infty}\sqrt{\rho_0 \over 2 \pi}\,\mrm{e}^{\mrm{i} \, r k_F x}  \mrm{e}^{-\mrm{i} \, (r \phi_0 -\theta_0)}
\end{align}
where $\mrm{U}_r$ is the appropriate Klein factor.
The density-density interaction $g$ and the pairing $u$ in the Hamiltonian (\ref{Matveev_Hamiltonian}) originate in the coulomb interaction.\cite{Meyer2009}

In order to decouple the pairing term in the Hamiltonian (\ref{Matveev_Hamiltonian}) the authors of Ref. [\onlinecite{Sitte2009}] apply the unitary transformation
\begin{align}
U = \mrm{e} ^{\mrm{i}\int \,dx' \, \theta_0 \psi_1 ^\dag \psi_1 }\,. \label{Unitary_non_local_Transformation}
\end{align}
such that the transformed operators are given by
\begin{align}
\tilde\psi_0 &= U^\dag \psi_0 U \sim \label{band_1_Fermi_operator}\\ &\sum_{r=R,L}\mrm{e}^{\mrm{i} \, r k_F x}  \mrm{e}^{-\mrm{i} \, (r \phi_0 -\theta_0)}\mrm{e}^{-r{\mrm{i} \,\pi \over 2} \int dx' \mrm{sign}(x-x') \psi_1 ^\dag \psi_1 } \nonumber
\end{align}
and
\begin{align}
\tilde\psi_1=U^\dag \psi_1 U = \mrm{e}^{-\mrm{i}\theta_0}\psi_1\,.\label{band_2_Fermi_operator}
\end{align}
The Hamiltonian (\ref{Matveev_Hamiltonian}) then assumes the form
\begin{align}
H = {u_0\over 2\pi}& \int dx\, \left[ K (\partial_x \theta_\rho) ^2 +{1\over K} (\partial_x \phi_\rho) ^2 \right]\label{Matveev_Hamiltonian_2} \\
 +&\int dx \, \xi  \left[ -\mrm{i}\,{u_1\over 2}\,\partial_x \tau^z  + 2 \,\d \,\tau^y \right]\xi \nonumber \\
 -&\int dx  \,{\lambda\over \pi}\, \partial_x \phi_\rho \psi_1 ^\dag \psi_1 \,.\nonumber
\end{align}
Here the curvature of the second sub-band has been neglected compared to the linear pairing term $u$. We have also switched the notations here such that $\partial_x\phi_\rho = \partial_x\phi_0 + \pi \psi_1 ^\dag \psi_1 $ is the fluctuation in the total density, and $\theta_\rho = \theta_0$ is the conjugate phase.
We would also like to note that one can recover the Majorana theory (\ref{Fermion_Hamiltonian}) by decomposing $\tilde\psi_1$ into it's real and imaginary parts as follows
\begin{align}
\tilde\psi_1 = {\mrm{e}^{\mrm{i} {\pi\over 4}}\over \sqrt{2}}\, \left( \xi_R  + \mrm{i}\, \xi_L  \right)\,.
\end{align}

%Aside from the coupling $\lambda = g_x - \pi v/K$
The Hamiltonian (\ref{Matveev_Hamiltonian_2}) and (\ref{H_rho}),(\ref{H_sigma}) and (\ref{Coupling_Term}) are identical. They both consist of two sectors a Luttinger liquid and single non-chiral massive Majorana field. Therefore, we identify our theory as the strong coupling limit of this one.

\begin{figure}
\centering
\includegraphics[width=7.6cm,height=5.5cm]{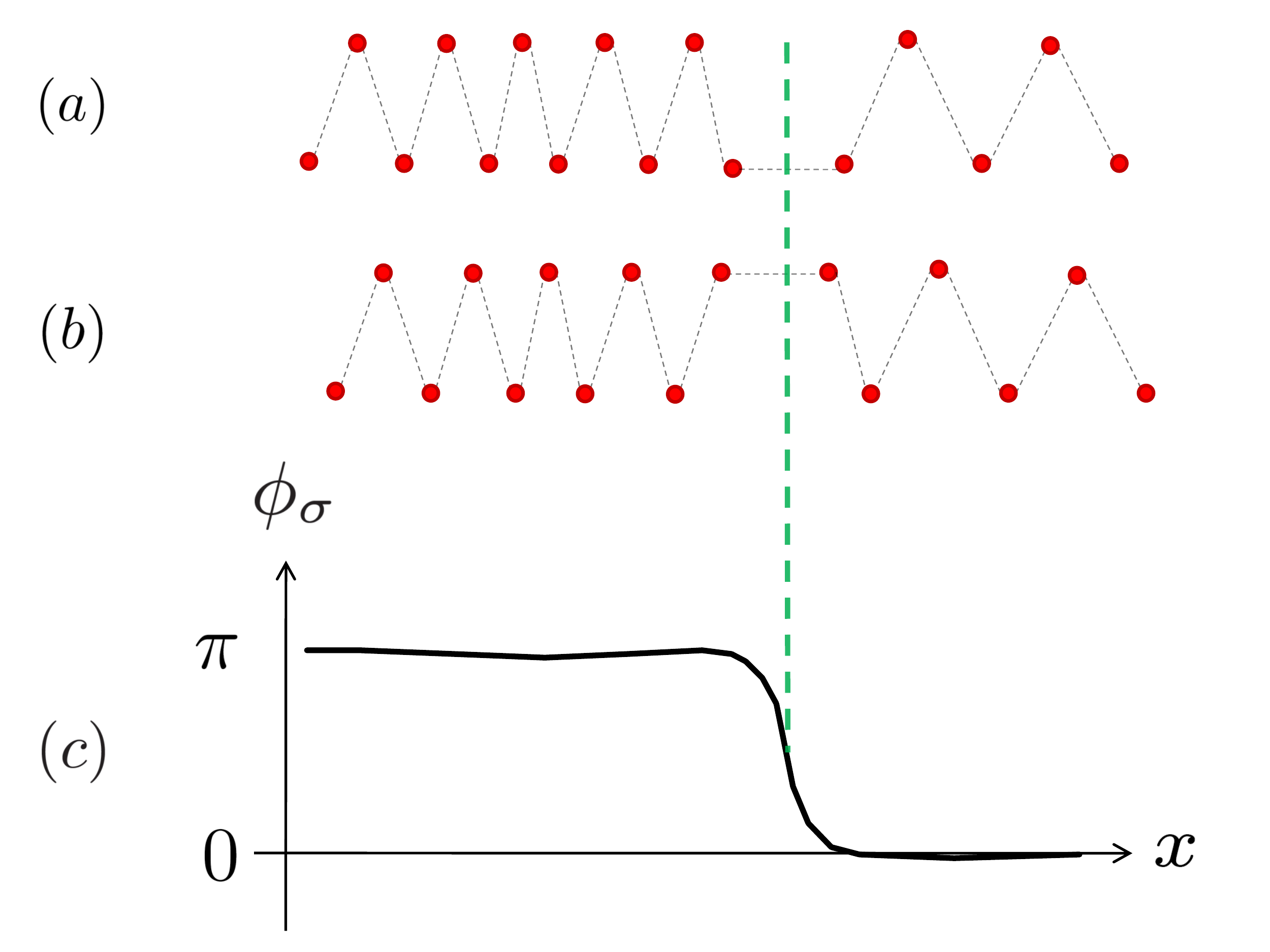}
\caption{(a) and (b) A chain with a soliton that is created by a defect in the $s=\uparrow$ and $s=\downarrow$ state respectively. (c) In both cases the field $\phi_\sigma$ is shifted by $-\pi$ due to the soliton.} \label{fig.1_Supp}
\end{figure}

Apparently, the expressions for the single particle operators (\ref{band_1_Fermi_operator}) and (\ref{band_2_Fermi_operator}) are sufficient to express the Fermion operators in the $s=\uparrow,\downarrow$ state in the zigzag. However, the operator (\ref{band_1_Fermi_operator}) is written in an inconveniently non-local form. We can overcome this inconvenience by replacing (\ref{band_1_Fermi_operator}) with a phenomenological expression. As shown in Fig. \ref{fig.1_Supp} the insertion of a particle both into the $s=\uparrow$ and $s=\downarrow$ states in the zigzag shifts the field $\phi_\sigma $ by $\pm\pi$. The operator that shifts $\phi_\sigma $ by $\pm\pi$ is $\mrm{e}^{\pm\mrm{i}\, \theta_\sigma}$. Therefore, we replace the non local string with $\theta_\sigma$
\begin{align}
\psi_0 \sim\sum_{r=R,L}\mrm{e}^{\mrm{i} \, r k_F x}  \mrm{e}^{-\mrm{i} \, (r \phi_\rho -\theta_\rho)}\mrm{e}^{r{\mrm{i} \, \theta_\sigma}}
\end{align}
This operator creates a soliton in the $\sigma$ sector in addition to a plasmonic excitation in the charge sector. Similarly, the operator $\psi_1$ has the form of a spinon multiplied by the factor $\mrm{e}^{\,\mrm{i}\, \theta_\rho}$ that inserts a charge at zero momentum.

We can now write expressions for second quantized Fermion operators inserted to the up and down positions in the tube $s=\uparrow,\downarrow$ as the appropriate superpositions of Fermions from the symmetric and anti symmetric sub-bands
\begin{align}
\Psi_{F,s} = {1\over \sqrt{2}}(\psi_0 + s \, \psi_1 ) \, . \label{Fermionic_Operator_2}
\end{align}

\section{Tunneling DOS on the edge}\label{Tunneling_DOS_appendix}

\subsection{Bosons}
The low energy limit of the second quantized Bose operator is (\ref{Bosonic_Operator_3})
\begin{align}
\Psi_{B,s}^\dag\simeq& \,\label{Bosonic_Operator_4} \b_0\,\sqrt{\rho_0}\, \mrm{e}^{-\mrm{i}\,(\theta_\rho + s\, \theta_\sigma)}+ \b_1\,\sqrt{2\pi \Delta_0 \over  u_\sigma}\mrm{e}^{-{2\Delta_0 \over u_\sigma}|x|}\,\, \times\\
 & \mrm{e}^{-\mrm{i}\,\theta_\rho } \bigg[  \mrm{e}^{\mrm{i}(\phi_\rho+2\pi\rho_0 x) } -\mrm{i}\, s \,\,\mrm{e}^{-\mrm{i}(\phi_\rho +2\pi\rho_0 x)} \bigg]\gamma_0 \, .\nonumber
\end{align}
The Bose operator is used to compute the single particle Green's function in the effective field theory (\ref{H_rho},\ref{H_sigma}). The contribution of the first term in the Bose operator is found by noting that the disorder parameter $e^{\pm i\t_\s}$ has an expectation value in the disordered side of the interface that penetrates to the ordered side and decays there as $e^{-x/l}$ ($l=\tu_\s/4\D_0$ is the correlation length in the ordered phase). We can therefore replace the operator $e^{\pm i\t_\s}$ by this expectation value. The contribution from the second component of the Bose operator is computed using the form of the zero-mode (\ref{Zero_Energy_State}) and the Luttinger liquid which describes the density modes. Putting all this together we have
\begin{align}
G_B&(\tau,x)=-\langle T_\tau \Psi_{B,s}(\tau,x) \Psi_{B,s} ^\dag (0,x) \rangle\label{Greens_function_Bosons} \\ \simeq&-|\b_0|^2\,\rho_0\sqrt{1\over 1+{ 2\rho_0\,l  }} {\mrm{e}^{-x/l}\over 1+\mrm{e}^{-x/l}}\left({1\over \rho_0  u_\rho|\tau|}\right)^{ { 1\over 4 K_\rho}}\nonumber\\&-|\b_1|^2\,{\pi \over 2l} \mrm{e}^{-|x|/l}\left({1\over \rho_0  u_\rho|\tau|}\right)^{K_\rho + { 1\over 4 K_\rho}}\nonumber\,.
\end{align}
where we have used the fact that
\begin{align}
\langle T_\tau \gamma_0 (\tau) \gamma_0(0)\rangle ={1\over2}\mrm{Sign}(\tau)\,.\label{fact_2}
\end{align}
To obtain the retarded greens function we analytically continue the function (\ref{Greens_function_Bosons})
\begin{align}
&G_B(t,x)= \\ &\simeq\Theta(t) \bigg[\sin {\pi \over 8 K_\rho}\sqrt{\rho_0 ^2\over 1+{2 \rho_0\,l }} {|\b_0|^2\,\mrm{e}^{-x/l}\over 1+\mrm{e}^{-x/l}}\left({1\over \rho_0  u_\rho t}\right)^{ { 1\over 4 K_\rho}}+\nonumber\\&{\pi  |\b_1|^2 \over 2l}\,\sin\left[{\pi \over 2}\left( {K_\rho}+{1 \over 4 K_\rho}\right)\right] \mrm{e}^{-|x|/l}\left({1\over \rho_0  u_\rho t}\right)^{K_\rho + { 1\over 4 K_\rho}}\bigg]\nonumber\,,
\end{align}
The imaginary part of the Fourier transformed version of this function gives the local DOS:
\begin{align}
&A_B(\omega,x) \simeq\label{Local_DOS_Bosons}  {B_0\, |\omega|^{ { 1\over 4 K_\rho}-1}\over 1+\mrm{e}^{\,x/l}}+B_1\,\mrm{e}^{|x|/l} |\omega|^{K_\rho + { 1\over 4 K_\rho}-1}\,.
\end{align}
where,
\[ B_0\simeq|\b_0|^2\,\rho_0\sqrt{\pi / 2 \over (1+{2\rho_0\,l })}{1\over \Gamma[1/4K_\rho]}\left({1\over\rho_0  u_\rho}\right)^{1\over 4K_\rho} \]
and
\[B_1 \simeq |\b_1|^2\,{1\over 2l} \,{\sqrt{\pi^3 / 8}\over \Gamma[K_\rho + { 1/ 4 K_\rho}]}\left({1\over \rho_0  u_\rho}\right)^{K_\rho + {1\over 4K_\rho}}\]
The leading contribution to the local DOS stems from the penetration of the disorder field into the ordered phase. The sub-leading contribution originates from the Majorana edge state.

\subsection{Fermions}
The low energy limit of the Fermion operator was found above and is given by
\begin{align}
\Psi_{F,s} \simeq{1\over \sqrt{2}} \bigg[\a_0\,\sum_{r=R,L}\sqrt{\rho_0}&\mrm{e}^{\mrm{i} \, r k_F x} \mrm{e}^{-\mrm{i} \, (r \phi_\rho -\theta_\rho)}\mrm{e}^{r{\mrm{i} \, \theta_\sigma}} \\+ &s\,\a_1\,\mrm{i}\,\sqrt{\Delta_0 \over 4 u_\sigma}\mrm{e}^{\mrm{i}\, \theta_\rho}\mrm{e}^{{u_\sigma \over \Delta_0}|x|}\gamma_0 \bigg]\,.\nonumber
\end{align}
We use this operator in the effective low energy theory (\ref{H_rho},\ref{H_sigma}) to obtain the imaginary time Green's function in the same way as discussed above for the Bosonic case. This gives
\begin{align}
G_F (\tau,x) \simeq - \mrm{Sign}&(\tau)\times \\  \bigg[{|\a_0|^2\rho_0\over\sqrt{ 1+{2 \rho_0 \,l}}}& {\mrm{e}^{-x/l}\over 1+\mrm{e}^{-x/l}}\left({1\over \rho_0  u_\rho|\tau|}\right)^{ K_\rho+{ 1\over 4 K_\rho}}\nonumber\\&+|\a_1|^2{\pi  \over 2 l} \mrm{e}^{-|x|/l}\left({1\over \rho_0 u_\rho |\tau|}\right)^{{ 1\over 4 K_\rho}}\bigg]\nonumber\,.
\end{align}
To obtain the low-energy local DOS we analytically continue this expression and the imaginary part in fourier space
\begin{align}
A_F(\omega,x) \simeq\label{Local_DOS_Fermions}  {A_0\, |\omega|^{ K_\rho+{ 1\over 4 K_\rho}-1} \over 1+\mrm{e}^{\,x/l}}  +A_1\,\mrm{e}^{|x|/l} |\omega|^{ { 1\over 4 K_\rho}-1}
\end{align}
where,
\[A_0 \simeq|\a_0|^2\rho_0\sqrt{1\over 1+{2\rho_0 \,l}}{\sqrt{\pi / 2}\over \Gamma[K_\rho+1/4K_\rho]}\left({1\over \rho_0  u_\rho}\right)^{K_\rho+{1\over 4K_\rho}}\]
and
\[A_1 \simeq|\a_1|^2{1\over 2l} \,{\sqrt{\pi^3 / 8}\over \Gamma[ { 1/ 4 K_\rho}]} \left({1\over \rho_0  u_\rho}\right)^{1\over 4K_\rho}\]
In this case the leading contribution comes from the localized Majorana Fermion, and the sub-leading one stems from the  penetration of the disorder parameter into the ordered phase.
\bibliographystyle{phd-url-notitle}

\begin{thebibliography}{29}
\expandafter\ifx\csname natexlab\endcsname\relax\def\natexlab#1{#1}\fi
\expandafter\ifx\csname bibnamefont\endcsname\relax
  \def\bibnamefont#1{#1}\fi
\expandafter\ifx\csname bibfnamefont\endcsname\relax
  \def\bibfnamefont#1{#1}\fi
\expandafter\ifx\csname citenamefont\endcsname\relax
  \def\citenamefont#1{#1}\fi
\expandafter\ifx\csname url\endcsname\relax
  \def\url#1{\texttt{#1}}\fi
\expandafter\ifx\csname urlprefix\endcsname\relax\def\urlprefix{URL }\fi
\providecommand{\bibinfo}[2]{#2}
\providecommand{\eprint}[2][]{\url{#2}}

\bibitem[{\citenamefont{Lahaye et~al.}(2007)\citenamefont{Lahaye, Koch,
  Fr\"{o}hlich, Fattori, Metz, Griesmaier, Giovanazzi, and Pfau}}]{Lahaye2007b}
\bibinfo{author}{\bibfnamefont{T.}~\bibnamefont{Lahaye}},
  \bibinfo{author}{\bibfnamefont{T.}~\bibnamefont{Koch}},
  \bibinfo{author}{\bibfnamefont{B.}~\bibnamefont{Fr\"{o}hlich}},
  \bibinfo{author}{\bibfnamefont{M.}~\bibnamefont{Fattori}},
  \bibinfo{author}{\bibfnamefont{J.}~\bibnamefont{Metz}},
  \bibinfo{author}{\bibfnamefont{A.}~\bibnamefont{Griesmaier}},
  \bibinfo{author}{\bibfnamefont{S.}~\bibnamefont{Giovanazzi}},
  \bibnamefont{and} \bibinfo{author}{\bibfnamefont{T.}~\bibnamefont{Pfau}},
  \bibinfo{journal}{Nature} \textbf{\bibinfo{volume}{448}},
  \bibinfo{pages}{672} (\bibinfo{year}{2007}),
  \href{http://www.ncbi.nlm.nih.gov/pubmed/17687319}{URL}.

\bibitem[{\citenamefont{Ni et~al.}(2008)\citenamefont{Ni, Ospelkaus,
  de~Miranda, Pe'er, Neyenhuis, Zirbel, Kotochigova, Julienne, Jin, and
  Ye}}]{Ni2008}
\bibinfo{author}{\bibfnamefont{K.-K.} \bibnamefont{Ni}},
  \bibinfo{author}{\bibfnamefont{S.}~\bibnamefont{Ospelkaus}},
  \bibinfo{author}{\bibfnamefont{M.~H.~G.} \bibnamefont{de~Miranda}},
  \bibinfo{author}{\bibfnamefont{A.}~\bibnamefont{Pe'er}},
  \bibinfo{author}{\bibfnamefont{B.}~\bibnamefont{Neyenhuis}},
  \bibinfo{author}{\bibfnamefont{J.~J.} \bibnamefont{Zirbel}},
  \bibinfo{author}{\bibfnamefont{S.}~\bibnamefont{Kotochigova}},
  \bibinfo{author}{\bibfnamefont{P.~S.} \bibnamefont{Julienne}},
  \bibinfo{author}{\bibfnamefont{D.~S.} \bibnamefont{Jin}}, \bibnamefont{and}
  \bibinfo{author}{\bibfnamefont{J.}~\bibnamefont{Ye}},
  \bibinfo{journal}{Science (New York, N.Y.)} \textbf{\bibinfo{volume}{322}},
  \bibinfo{pages}{231} (\bibinfo{year}{2008}),
  \href{http://www.ncbi.nlm.nih.gov/pubmed/18801969}{URL}.

\bibitem[{\citenamefont{Pupillo et~al.}(2010)\citenamefont{Pupillo, Micheli,
  Boninsegni, Lesanovsky, and Zoller}}]{Pupillo2010}
\bibinfo{author}{\bibfnamefont{G.}~\bibnamefont{Pupillo}},
  \bibinfo{author}{\bibfnamefont{A.}~\bibnamefont{Micheli}},
  \bibinfo{author}{\bibfnamefont{M.}~\bibnamefont{Boninsegni}},
  \bibinfo{author}{\bibfnamefont{I.}~\bibnamefont{Lesanovsky}},
  \bibnamefont{and} \bibinfo{author}{\bibfnamefont{P.}~\bibnamefont{Zoller}},
  \bibinfo{journal}{Physical Review Letters} \textbf{\bibinfo{volume}{104}},
  \bibinfo{pages}{223002} (\bibinfo{year}{2010}),
  \href{http://link.aps.org/doi/10.1103/PhysRevLett.104.223002}{URL}.

\bibitem[{\citenamefont{{Dalla Torre} et~al.}(2006)\citenamefont{{Dalla Torre},
  Berg, and Altman}}]{HIprl}
\bibinfo{author}{\bibfnamefont{E.~G.} \bibnamefont{{Dalla Torre}}},
  \bibinfo{author}{\bibfnamefont{E.}~\bibnamefont{Berg}}, \bibnamefont{and}
  \bibinfo{author}{\bibfnamefont{E.}~\bibnamefont{Altman}},
  \bibinfo{journal}{Physical Review Letters} \textbf{\bibinfo{volume}{97}},
  \bibinfo{pages}{260401} (\bibinfo{year}{2006}),
  \href{http://link.aps.org/doi/10.1103/PhysRevLett.97.260401}{URL}.

\bibitem[{\citenamefont{Cooper and Shlyapnikov}(2009)}]{Cooper2009}
\bibinfo{author}{\bibfnamefont{N.~R.} \bibnamefont{Cooper}} \bibnamefont{and}
  \bibinfo{author}{\bibfnamefont{G.~V.} \bibnamefont{Shlyapnikov}},
  \bibinfo{journal}{Physical Review Letters} \textbf{\bibinfo{volume}{103}},
  \bibinfo{pages}{155302} (\bibinfo{year}{2009}),
  \href{http://link.aps.org/doi/10.1103/PhysRevLett.103.155302}{URL}.

\bibitem[{\citenamefont{Fishman et~al.}(2008)\citenamefont{Fishman, {De
  Chiara}, Calarco, and Morigi}}]{Fishman2008}
\bibinfo{author}{\bibfnamefont{S.}~\bibnamefont{Fishman}},
  \bibinfo{author}{\bibfnamefont{G.}~\bibnamefont{{De Chiara}}},
  \bibinfo{author}{\bibfnamefont{T.}~\bibnamefont{Calarco}}, \bibnamefont{and}
  \bibinfo{author}{\bibfnamefont{G.}~\bibnamefont{Morigi}},
  \bibinfo{journal}{Physical Review B} \textbf{\bibinfo{volume}{77}},
  \bibinfo{pages}{064111} (\bibinfo{year}{2008}),
  \href{http://link.aps.org/doi/10.1103/PhysRevB.77.064111}{URL}.

\bibitem[{\citenamefont{Birkl et~al.}(1992)\citenamefont{Birkl, Kassner, and
  Walther}}]{Birkl1992}
\bibinfo{author}{\bibfnamefont{G.}~\bibnamefont{Birkl}},
  \bibinfo{author}{\bibfnamefont{S.}~\bibnamefont{Kassner}}, \bibnamefont{and}
  \bibinfo{author}{\bibfnamefont{H.}~\bibnamefont{Walther}},
  \bibinfo{journal}{Nature} \textbf{\bibinfo{volume}{357}},
  \bibinfo{pages}{310} (\bibinfo{year}{1992}),
  \href{http://www.nature.com/doifinder/10.1038/357310a0}{URL}.

\bibitem[{\citenamefont{Shimshoni
  et~al.}(2011{\natexlab{a}})\citenamefont{Shimshoni, Morigi, and
  Fishman}}]{Shimshoni2011}
\bibinfo{author}{\bibfnamefont{E.}~\bibnamefont{Shimshoni}},
  \bibinfo{author}{\bibfnamefont{G.}~\bibnamefont{Morigi}}, \bibnamefont{and}
  \bibinfo{author}{\bibfnamefont{S.}~\bibnamefont{Fishman}},
  \bibinfo{journal}{Physical Review Letters} \textbf{\bibinfo{volume}{106}},
  \bibinfo{pages}{010401} (\bibinfo{year}{2011}{\natexlab{a}}),
  \href{http://link.aps.org/doi/10.1103/PhysRevLett.106.010401}{URL}.

\bibitem[{\citenamefont{Shimshoni
  et~al.}(2011{\natexlab{b}})\citenamefont{Shimshoni, Morigi, and
  Fishman}}]{Shimshoni2011a}
\bibinfo{author}{\bibfnamefont{E.}~\bibnamefont{Shimshoni}},
  \bibinfo{author}{\bibfnamefont{G.}~\bibnamefont{Morigi}}, \bibnamefont{and}
  \bibinfo{author}{\bibfnamefont{S.}~\bibnamefont{Fishman}},
  \bibinfo{journal}{Physical Review A} \textbf{\bibinfo{volume}{83}},
  \bibinfo{pages}{032308} (\bibinfo{year}{2011}{\natexlab{b}}),
  \href{http://link.aps.org/doi/10.1103/PhysRevA.83.032308}{URL}.

\bibitem[{\citenamefont{Meyer et~al.}(2007)\citenamefont{Meyer, Matveev, and
  Larkin}}]{Meyer2007}
\bibinfo{author}{\bibfnamefont{J.~S.} \bibnamefont{Meyer}},
  \bibinfo{author}{\bibfnamefont{K.~A.} \bibnamefont{Matveev}},
  \bibnamefont{and} \bibinfo{author}{\bibfnamefont{A.~I.}~\bibnamefont{Larkin}},
  \bibinfo{journal}{Physical Review Letters} \textbf{\bibinfo{volume}{98}},
  \bibinfo{pages}{126404} (\bibinfo{year}{2007}),
  \href{http://link.aps.org/doi/10.1103/PhysRevLett.98.126404}{URL}.

\bibitem[{\citenamefont{Meyer and Matveev}(2009)}]{Meyer2009}
\bibinfo{author}{\bibfnamefont{J.~S.} \bibnamefont{Meyer}} \bibnamefont{and}
  \bibinfo{author}{\bibfnamefont{K.~A.} \bibnamefont{Matveev}},
  \bibinfo{journal}{Journal of Physics: Condensed Matter}
  \textbf{\bibinfo{volume}{21}}, \bibinfo{pages}{023203}
  (\bibinfo{year}{2009}),
  \href{http://stacks.iop.org/0953-8984/21/i=2/a=023203?key=crossref.04c619cb37173dbcf2f5842ea92251c5}{URL}.

\bibitem[{\citenamefont{Sitte et~al.}(2009)\citenamefont{Sitte, Rosch, Meyer,
  Matveev, and Garst}}]{Sitte2009}
\bibinfo{author}{\bibfnamefont{M.}~\bibnamefont{Sitte}},
  \bibinfo{author}{\bibfnamefont{A.}~\bibnamefont{Rosch}},
  \bibinfo{author}{\bibfnamefont{J.~S.} \bibnamefont{Meyer}},
  \bibinfo{author}{\bibfnamefont{K.~A.} \bibnamefont{Matveev}},
  \bibnamefont{and} \bibinfo{author}{\bibfnamefont{M.}~\bibnamefont{Garst}},
  \bibinfo{journal}{Physical Review Letters} \textbf{\bibinfo{volume}{102}},
  \bibinfo{pages}{176404} (\bibinfo{year}{2009}),
  \href{http://link.aps.org/doi/10.1103/PhysRevLett.102.176404}{URL}.

\bibitem[{\citenamefont{Orignac and Giamarchi}(1998)}]{Orignac1998}
\bibinfo{author}{\bibfnamefont{E.}~\bibnamefont{Orignac}} \bibnamefont{and}
  \bibinfo{author}{\bibfnamefont{T.}~\bibnamefont{Giamarchi}},
  \bibinfo{journal}{Physical Review B} \textbf{\bibinfo{volume}{57}},
  \bibinfo{pages}{11713} (\bibinfo{year}{1998}),
  \href{http://link.aps.org/doi/10.1103/PhysRevB.57.11713}{URL}.

\bibitem[{\citenamefont{Starykh et~al.}(2000)\citenamefont{Starykh, Maslov,
  H\"ausler, and Glazman}}]{Starykh2000}
\bibinfo{author}{\bibfnamefont{O.}~\bibnamefont{Starykh}},
  \bibinfo{author}{\bibfnamefont{D.}~\bibnamefont{Maslov}},
  \bibinfo{author}{\bibfnamefont{W.}~\bibnamefont{H\"ausler}},
  \bibnamefont{and} \bibinfo{author}{\bibfnamefont{L.}~\bibnamefont{Glazman}},
  \bibinfo{journal}{Low-Dimensional Systems: Interactions and Transport
  Properties} pp. \bibinfo{pages}{37--78} (\bibinfo{year}{2000}),
  \href{http://www.springerlink.com/index/34370U4702832656.pdf}{URL}.

\bibitem[{\citenamefont{Astrakharchik et~al.}(2009)\citenamefont{Astrakharchik,
  {De Chiara}, Morigi, and Boronat}}]{Astrakharchik2009}
\bibinfo{author}{\bibfnamefont{G.~E.} \bibnamefont{Astrakharchik}},
  \bibinfo{author}{\bibfnamefont{G.}~\bibnamefont{{De Chiara}}},
  \bibinfo{author}{\bibfnamefont{G.}~\bibnamefont{Morigi}}, \bibnamefont{and}
  \bibinfo{author}{\bibfnamefont{J.}~\bibnamefont{Boronat}},
  \bibinfo{journal}{Journal of Physics B: Atomic, Molecular and Optical
  Physics} \textbf{\bibinfo{volume}{42}}, \bibinfo{pages}{154026}
  (\bibinfo{year}{2009}),
  \href{http://stacks.iop.org/0953-4075/42/i=15/a=154026?key=crossref.1b4d5f8bca7fde35ce67d19607f7947e}{URL}.

\bibitem[{\citenamefont{Matveev}(2004)}]{Matveev2004}
\bibinfo{author}{\bibfnamefont{K.~A.} \bibnamefont{Matveev}},
  \bibinfo{journal}{Physical Review B} \textbf{\bibinfo{volume}{70}},
  \bibinfo{pages}{245319} (\bibinfo{year}{2004}),
  \href{http://link.aps.org/doi/10.1103/PhysRevB.70.245319}{URL}.

\bibitem[{\citenamefont{Shelton et~al.}(1996)\citenamefont{Shelton, Nersesyan,
  and Tsvelik}}]{Shelton1996}
\bibinfo{author}{\bibfnamefont{D.~G.} \bibnamefont{Shelton}},
  \bibinfo{author}{\bibfnamefont{A.~A.} \bibnamefont{Nersesyan}},
  \bibnamefont{and} \bibinfo{author}{\bibfnamefont{A.~M.}
  \bibnamefont{Tsvelik}}, \bibinfo{journal}{Physical review. B, Condensed
  matter} \textbf{\bibinfo{volume}{53}}, \bibinfo{pages}{8521}
  (\bibinfo{year}{1996}), ISSN \bibinfo{issn}{0163-1829},
  \href{http://www.ncbi.nlm.nih.gov/pubmed/9982358}{URL}.

\bibitem[{\citenamefont{Lecheminant et~al.}(2002)\citenamefont{Lecheminant,
  Gogolin, and Nersesyan}}]{Lecheminant2002}
\bibinfo{author}{\bibfnamefont{P.}~\bibnamefont{Lecheminant}},
  \bibinfo{author}{\bibfnamefont{A.~O.} \bibnamefont{Gogolin}},
  \bibnamefont{and} \bibinfo{author}{\bibfnamefont{A.~A.}
  \bibnamefont{Nersesyan}}, \bibinfo{journal}{Nuclear Physics B}
  \textbf{\bibinfo{volume}{639}}, \bibinfo{pages}{502} (\bibinfo{year}{2002}),
  \href{http://linkinghub.elsevier.com/retrieve/pii/S0550321302004741}{URL}.

\bibitem[{\citenamefont{Zuber and Itzykson}(1977)}]{Zuber1977}
\bibinfo{author}{\bibfnamefont{J. ~B.}~\bibnamefont{Zuber}} \bibnamefont{and}
  \bibinfo{author}{\bibfnamefont{C.}~\bibnamefont{Itzykson}},
  \bibinfo{journal}{Physical Review D} \textbf{\bibinfo{volume}{15}},
  \bibinfo{pages}{2875} (\bibinfo{year}{1977}),
  \href{http://link.aps.org/doi/10.1103/PhysRevD.15.2875}{URL}.

\bibitem[{\citenamefont{Kitaev}(2001)}]{Kitaev2001}
\bibinfo{author}{\bibfnamefont{A.~Y.} \bibnamefont{Kitaev}},
  \bibinfo{journal}{Phys.-Usp.} \textbf{\bibinfo{volume}{44}},
  \bibinfo{pages}{131} (\bibinfo{year}{2001}),
  \href{http://iopscience.iop.org/1063-7869/44/10S/S29}{URL}.

\bibitem[{\citenamefont{Kollath et~al.}(2007)\citenamefont{Kollath, K\"{o}hl,
  and Giamarchi}}]{Kollath2007b}
\bibinfo{author}{\bibfnamefont{C.}~\bibnamefont{Kollath}},
  \bibinfo{author}{\bibfnamefont{M.}~\bibnamefont{K\"{o}hl}}, \bibnamefont{and}
  \bibinfo{author}{\bibfnamefont{T.}~\bibnamefont{Giamarchi}},
  \bibinfo{journal}{Physical Review A} \textbf{\bibinfo{volume}{76}},
  \bibinfo{pages}{063602} (\bibinfo{year}{2007}),
  \href{http://link.aps.org/doi/10.1103/PhysRevA.76.063602}{URL}.

\bibitem[{\citenamefont{Bakr et~al.}(2010)\citenamefont{Bakr, Peng, Tai, Ma,
  Simon, Gillen, F\"{o}lling, Pollet, and Greiner}}]{Bakr2010}
\bibinfo{author}{\bibfnamefont{W.~S.} \bibnamefont{Bakr}},
  \bibinfo{author}{\bibfnamefont{A.}~\bibnamefont{Peng}},
  \bibinfo{author}{\bibfnamefont{M.~E.} \bibnamefont{Tai}},
  \bibinfo{author}{\bibfnamefont{R.}~\bibnamefont{Ma}},
  \bibinfo{author}{\bibfnamefont{J.}~\bibnamefont{Simon}},
  \bibinfo{author}{\bibfnamefont{J.~I.} \bibnamefont{Gillen}},
  \bibinfo{author}{\bibfnamefont{S.}~\bibnamefont{F\"{o}lling}},
  \bibinfo{author}{\bibfnamefont{L.}~\bibnamefont{Pollet}}, \bibnamefont{and}
  \bibinfo{author}{\bibfnamefont{M.}~\bibnamefont{Greiner}},
  \bibinfo{journal}{Science (New York, N.Y.)} \textbf{\bibinfo{volume}{329}},
  \bibinfo{pages}{547} (\bibinfo{year}{2010}),
  \href{http://www.ncbi.nlm.nih.gov/pubmed/20558666}{URL}.

\bibitem[{\citenamefont{Sherson et~al.}(2010)\citenamefont{Sherson, Weitenberg,
  Endres, Cheneau, Bloch, and Kuhr}}]{Sherson2010}
\bibinfo{author}{\bibfnamefont{J.~F.} \bibnamefont{Sherson}},
  \bibinfo{author}{\bibfnamefont{C.}~\bibnamefont{Weitenberg}},
  \bibinfo{author}{\bibfnamefont{M.}~\bibnamefont{Endres}},
  \bibinfo{author}{\bibfnamefont{M.}~\bibnamefont{Cheneau}},
  \bibinfo{author}{\bibfnamefont{I.}~\bibnamefont{Bloch}}, \bibnamefont{and}
  \bibinfo{author}{\bibfnamefont{S.}~\bibnamefont{Kuhr}},
  \bibinfo{journal}{Nature} \textbf{\bibinfo{volume}{467}}, \bibinfo{pages}{68}
  (\bibinfo{year}{2010}),
  \href{http://www.ncbi.nlm.nih.gov/pubmed/20720540}{URL}.

\bibitem[{\citenamefont{White}(1992)}]{White1992}
\bibinfo{author}{\bibfnamefont{S.~R.} \bibnamefont{White}},
  \bibinfo{journal}{Physical Review Letters} \textbf{\bibinfo{volume}{69}},
  \bibinfo{pages}{2863} (\bibinfo{year}{1992}),
  \href{http://link.aps.org/doi/10.1103/PhysRevLett.69.2863}{URL}.

\bibitem[{\citenamefont{Endres et~al.}(2011)\citenamefont{Endres, Cheneau,
  Fukuhara, Weitenberg, Schau\ss{}, Gross, Mazza, Ba\"nuls, Pollet, Bloch
  et~al.}}]{Endres2011}
\bibinfo{author}{\bibfnamefont{M.}~\bibnamefont{Endres}},
  \bibinfo{author}{\bibfnamefont{M.}~\bibnamefont{Cheneau}},
  \bibinfo{author}{\bibfnamefont{T.}~\bibnamefont{Fukuhara}},
  \bibinfo{author}{\bibfnamefont{C.}~\bibnamefont{Weitenberg}},
  \bibinfo{author}{\bibfnamefont{P.}~\bibnamefont{Schau\ss{}}},
  \bibinfo{author}{\bibfnamefont{C.}~\bibnamefont{Gross}},
  \bibinfo{author}{\bibfnamefont{L.}~\bibnamefont{Mazza}},
  \bibinfo{author}{\bibfnamefont{M.~C.} \bibnamefont{Ba\"nuls}},
  \bibinfo{author}{\bibfnamefont{L.}~\bibnamefont{Pollet}},
  \bibinfo{author}{\bibfnamefont{I.}~\bibnamefont{Bloch}},
  \bibnamefont{et~al.}, \bibinfo{journal}{Science}
  \textbf{\bibinfo{volume}{334}}, \bibinfo{pages}{200} (\bibinfo{year}{2011}),
  \href{http://www.sciencemag.org/content/334/6053/200.abstract}{URL}.

\bibitem[{\citenamefont{Kestner et~al.}(2011)\citenamefont{Kestner, Wang, Sau,
  and Sarma}}]{Kestner2011}
\bibinfo{author}{\bibfnamefont{J.~P.} \bibnamefont{Kestner}},
  \bibinfo{author}{\bibfnamefont{B.}~\bibnamefont{Wang}},
  \bibinfo{author}{\bibfnamefont{J.~D.} \bibnamefont{Sau}}, \bibnamefont{and}
  \bibinfo{author}{\bibfnamefont{S.} \bibnamefont{DasSarma}},
  \bibinfo{journal}{Physical Review B} \textbf{\bibinfo{volume}{83}},
  \bibinfo{pages}{174409} (\bibinfo{year}{2011}),
  \href{http://link.aps.org/doi/10.1103/PhysRevB.83.174409}{URL}.

\bibitem[{Bos()}]{Bosonization_Books}
\bibinfo{note}{Readers who seek general information regarding the
  re-fermionization procedure may consult e.g. Ref.\onlinecite{GiamarchiBook}
  or Ref.\onlinecite{GogolinBook}.}

\bibitem[{\citenamefont{Giamarchi}(2003)}]{GiamarchiBook}
\bibinfo{author}{\bibfnamefont{T.}~\bibnamefont{Giamarchi}},
  \emph{\bibinfo{title}{Quantum Phyiscs in One Dimension}}
  (\bibinfo{publisher}{Oxford Sceince Publications}, \bibinfo{address}{Oxford},
  \bibinfo{year}{2003}).

\bibitem[{\citenamefont{Gogoglin et~al.}(1998)\citenamefont{Gogoglin,
  Nersesyan, and Tsvelik}}]{GogolinBook}
\bibinfo{author}{\bibfnamefont{A.~O.} \bibnamefont{Gogoglin}},
  \bibinfo{author}{\bibfnamefont{A.~A.} \bibnamefont{Nersesyan}},
  \bibnamefont{and} \bibinfo{author}{\bibfnamefont{A.~M.}
  \bibnamefont{Tsvelik}}, \emph{\bibinfo{title}{Bosonization Approach to
  Strongly Correlated Systems}} (\bibinfo{publisher}{Cambridge University
  Press}, \bibinfo{address}{Cambridge}, \bibinfo{year}{1998}).

\end{thebibliography}

\end{document}